\documentclass[aps,prd,preprint,superscriptaddress,nofootinbib,showpacs]{revtex4}
\pdfoutput=1

\usepackage{slashed}
\usepackage{rotating}
\usepackage{color}
\usepackage{epsfig,latexsym,cancel,amssymb,amsmath,verbatim,mathrsfs}
\usepackage{color}
\usepackage{graphicx}

\newcommand{\wt}{\widetilde}

\newcommand{\gev}{\ensuremath{\,\mathrm{GeV}}}

\def\ra{\rightarrow}
\def\L{\left(}
\def\R{\right)}
\def\wt{\widetilde}

\def\ld{\lambda}
\def\f{\frac}
\newcommand{\be}{\begin{equation}}
\newcommand{\ee}{\end{equation}}
\newcommand{\bea}{\begin{eqnarray}}
\newcommand{\eea}{\end{eqnarray}}
\newcommand{\ba}{\begin{array}}
\newcommand{\ea}{\end{array}}

\newcommand\gmtwo{\ensuremath{(g-2)_\mu}}

\long\def\symbolfootnote[#1]#2{\begingroup%
\def\thefootnote{\fnsymbol{footnote}}\footnote[#1]{#2}\endgroup}

\newcommand{\beq}{\begin{equation}}
\newcommand{\eeq}{\end{equation}}

\begin{document}

\def\thefootnote{\fnsymbol{footnote}}

{\small
\begin{flushright}
KIAS-P15024,
IPMU15-0121\\
\end{flushright} }

\medskip

\begin{center}
{\bf
{\Large
LHC $\tau-$rich Tests of Lepton-specific 2HDM for $(g-2)_\mu$} }
\end{center}

\smallskip

\begin{center}{\large
Eung Jin Chun$^{a}$,
Zhaofeng Kang$^{a}$,
Michihisa Takeuchi$^{b}$,\\
Yue-Lin Sming Tsai$^{b}$
}
\end{center}

\begin{center}
{\em $^a$ School of Physics, Korea Institute for Advanced Study,
Seoul 130-722, Korea}\\[0.2cm]
{\em $^b$ Kavli IPMU (WPI), The University of Tokyo,
5-1-5 Kashiwanoha, Kashiwa, Chiba 277-8583, Japan}\\[0.2cm]
(\today)
\end{center}

\bigskip 

\centerline{\bf ABSTRACT}

{The lepton-sepcific (or type X) 2HDM (L2HDM) is an attractive new physics candidate explaining the muon  $g-2$ anomaly requiring a light CP-odd boson $A$ and large $\tan\beta$.  This scenario leads to  $\tau$-rich signatures, such as $3\tau$, $4\tau$ and $4\tau+W/Z$, which can be readily accessible at the LHC.
We first study the whole L2HDM parameter space 
to identify allowed regions of extra Higgs boson masses as well as two couplings $\lambda_{hAA}$ and $\xi_h^l$ which determine  the 125 GeV Higgs boson decays $h\to \tau^+\tau^-$ and $h\to AA/AA^*(\tau^+\tau^-)$, respectively. This motivates us to set up two regions of interest:  (A) $m_A \ll m_{H} \sim m_{H^\pm}$, and (B) $m_A \sim m_{H^\pm} \sim {\cal O}(100) \mbox{GeV} \ll m_H$,
for which derive the current constraints by adopting the chargino-neutralino search at the LHC8, and then analyze the LHC14 prospects by implementing $\tau$-tagging algorithm. A correlated study of the upcoming precision determination of the 125 GeV Higgs boson decay properties as well as the observation of multi-tau events at the next runs of LHC will be able to shed light on the L2HDM option for the muon $g-2$.}

\vspace{0.3in}


\newpage

\section{Introduction}

The muon {$g-2$} anomaly has been a long standing puzzle since the announcement by the E821 experiment in 2001 \cite{bnl}. During the past 15 years, development in both experimental and theoretical sides has been made to reduce the uncertainties by a factor of two or so, and thus establish a consistent $3 \sigma$ discrepancy.  Although not significant enough, it could be a sign of new physics beyond the Standard Model (SM). Since the first announcement of the muon {$g-2$} anomaly, quite a few studies have been made in the context of two-Higgs-doublets models (2HDMs)~\cite{dedes01,maria02,cheung03,cao09}.
Recently, it was realized that the ``lepton-specific" (or ``type X") 2HDM (L2HDM)\footnote{In the scale invariant 2HDM with one Higgs doublet triggering electroweak symmetry breaking, 
the heavy Higgs bosons should be around 400 GeV~\cite{2HDMSI}, which is excluded in the type-II but not in the type-X.} 
with a light CP-odd Higgs boson $A$ and large $\tan\beta$  is a promising candidate accommodating a large muon {$g-2$} while escaping all the existing theoretical and experimental constraints \cite{broggio14}.
Some of the following studies showed that the allowed parameter space is
further resrticted, in particular, by the consideration of the 125 GeV Higgs boson decay
to light CP-odd Higgs bosons $h\to AA$ \cite{wang14}, and the tau decay $\tau \to \mu \nu \nu$ combined with the lepton universality conditions \cite{abe15}.

In this paper, we attempt to make a thorough study of the whole L2HDM parameter 
space in favor of the muon {$g-2$} explanation, and analyze the LHC tests of the favoured 
parameter space leading to $\tau$-rich signatures like $3\tau$, $4\tau$ and $4\tau+W/Z$.
First, we show how the  SM Higgs exotic decays $h \to AA$ as well as  $h\to A A^*(\tau^+ \tau^-)$ 
constrain the parameter space.  It is connected to the determination of the allowed ranges 
of the normalized tau (lepton) Yukawa coupling in the right- or wrong-sign domain,
 and thus more precise measurement of the 125 GeV Higgs boson properties will 
 put stronger bounds on the L2HDM parameter space.  
As we will see, the $hAA$ coupling can be made arbitrarily small by a cancellation 
for $m_H \gg m_A$ only in the wrong-sign limit of the tau Yukawa coupling \cite{ferreira14}, and it opens 
up the region of $m_A < m_h/2$~\cite{wang14}. 
In the region of $m_A>m_h/2$, the three-body decay $h \to A \tau^+ \tau^-$ 
should be suppressed and the SM (right-sign) limit of the tau 
Yukawa coupling is allowed for $m_A \gtrsim 70$ GeV.  The allowed parameter space is further restircited by the lepton universality tests of HFAG which measures the leptonic decay processes at the level of 0.1\%~\cite{hfag14}.  For this, we improve the analysis of~\cite{abe15} to single out proper constraints on the tree and loop contributions to the tau decay.

After scanning the L2HDM parameter space, we identify two allowed regions: 
(A) the well-known region of  $m_A \ll m_{H} \sim m_{H^\pm}$ and 
(B) $m_A \sim m_{H^\pm} \sim {\cal O}(100) \mbox{GeV} \ll m_H$.
Most of these parameter regions predict $\tau$-rich signatures easily accessible at the LHC, 
and thus can be readily probed. As a first step, we investigate how the current LHC 8 TeV data 
constrain the two regions, and show that the most stringent constraint
comes from the chargino-neutralino searches. 
We found that the region (B) has already been excluded at 95~\% CL.
For the region (A), most of the allowed L2HDM parameter region can be probed soon at the next runs of LHC.

The paper is organized as follows.
In Section~\ref{sec:L2HDM}, we introduce the L2HDM to provide useful 
formulas, and explain why a large $(g-2)_\mu $ is easily accommodated 
with a light CP-odd Higgs boson $A$ and large $\tan\beta$. 
In Section~\ref{sec:scan}, we summarize all the relevant theoretical and experimental 
constraints, and quote some of the latest results which are not included in our analysis.
By using the profile likelihood method, 
we identify the allowed L2HDM parameter regions 
under these constraints and show them at 68\% and 95\% confidence level.  
In Section~\ref{sec:LHC}, we discuss $\tau$-rich signatures at the LHC expected in the identified parameter regions. 
We analyze the 3$\tau$ events to identify the parameter regions excluded already by the current LHC 8~TeV data. 
In addition, we show the prospect for the future LHC14 run with a dedicated simulation.
We conclude in Section~\ref{sec:conclusion}.

\section{2HDM with a lepton-specific doublet (L2HDM)}
\label{sec:L2HDM}

Let us first introduce the L2HDM to present useful formulas for our analysis heavily relying on the paper by Gunion and Haber \cite{2hdms}. Among various types of 2HDMs classified by the Yukawa coupling patterns of the two Higgs doublets $\Phi_{1,2}$ with the same SM quantum numbers, the L2HDM allows the following Yukawa couplings:
\begin{equation}
-{\cal L}_Y=Y^u\overline{ Q_L} \wt \Phi_2 u_R + Y^d  \overline{ Q_L} \Phi_2 d_R+Y^e\overline{ l_L} \Phi_1 e_R+c.c.,
\end{equation}
where family indices have been omitted and  $\wt \Phi_2=i\sigma_2\Phi_2^*$. This pattern may be a result of a discrete $Z_2$~\cite{softZ2}: $\Phi_2\ra \Phi_2$ and $\Phi_1\ra-\Phi_1$ combined with $e_R\ra -e_R$ while the other fermions are invariant under the $Z_2$ transformation. The most general form of the 2HDM scalar potential is given by
\begin{eqnarray}
\nonumber V_{\mathrm{2HDM}} &=& m_{11}^2\Phi_1^{\dagger}\Phi_1 + m_{22}^2\Phi_2^{\dagger}\Phi_2 -\Big[m_{12}^2\Phi_1^{\dagger}\Phi_2 + \mathrm{h.c.}\Big]
+\frac{1}{2}\lambda_1\left(\Phi_1^\dagger\Phi_1\right)^2+\frac{1}{2}\lambda_2\left(\Phi_2^\dagger\Phi_2\right)^2 \\
\nonumber && +\lambda_3\left(\Phi_1^\dagger\Phi_1\right)\left(\Phi_2^\dagger\Phi_2\right)+\lambda_4\left(\Phi_1^\dagger\Phi_2\right)\left(\Phi_2^\dagger\Phi_1\right)
+\Big\{ \frac{1}{2}\lambda_5\left(\Phi_1^\dagger\Phi_2\right)^2+\Big[\lambda_6\left(\Phi_1^\dagger\Phi_1\right) \\
&& +\lambda_7\left(\Phi_2^\dagger\Phi_2\right)\Big]\left(\Phi_1^\dagger\Phi_2\right) + \rm{h.c.}\Big\}.
\label{eq:2hdmgen}
\end{eqnarray}
The $Z_2$ symmetry enforces $\lambda_6=\lambda_7=0$. However, the $m_{12}^2$ term that softly breaks $Z_2$ should be allowed.  All couplings are assumed to be real. In the desired vacuum, both doublets acquire VEVs, denoted as $v_1$ and $v_2$ for $\Phi_1$ and $\Phi_2$, respectively. Large VEV hierarchy, i.e., $ \tan{\beta}\equiv v_2/v_1\gg 1$, is of our interest for the explanation of the muon {$g-2$}.

By decomposing the doublets as $\Phi_i=(H_i^+,(v_i+h_i+iA_i)/\sqrt{2})^T$,
we see the model has three mass squared matrices of $A_i$, $H_i^\pm$ and $h_i$, which can be diagonalized by two angles $\alpha$ and $\beta$. The physical Higgs particles in mass eigenstates are given by
\begin{eqnarray}\label{}
A&=&-s_\beta A_1+c_\beta A_2,\quad H^+=-s_\beta H_1^++c_\beta H^+_2,\cr
h&=&-s_\alpha h_1+c_\alpha h_2,\quad~ H=c_\alpha h_1+s_\alpha h_2,
\end{eqnarray}
where $s_\alpha$ and $s_\beta$ are abbreviations for sin$\,\alpha$ and sin$\,\beta$, etc.
In this paper, we adopt the convention $0<\beta<\pi/2$ and $-\pi/2\leq \beta-\alpha\leq\pi/2$. Then
the SM-like Higgs boson is  $h\approx c_\alpha h_2$ with either positive or negative sign for $c_\alpha$.
In the very large $ \tan\beta$ limit, two Higgs doublets are almost decoupled. But some degree of non-decoupling effects, 
encoded in $0\leq c_{\beta-\alpha}\ll1$, will play very important roles in our study.

The mass spectrum can be calculated analytically in terms of the coupling constants in the Higgs potential, but practically it is more convenient to take masses as inputs and inversely express coupling constants with them:
\begin{eqnarray}
\nonumber && \lambda_1 = \frac{m_H^2c_\alpha^2+m_h^2s_\alpha^2-m_{12}^2\tan\beta}{v^2c_\beta^2},\\
\nonumber && \lambda_2 = \frac{m_H^2s_\alpha^2+m_h^2c_\alpha^2-m_{12}^2\cot\beta}{v^2s_\beta^2},\\
\nonumber && \lambda_3 = \frac{(m_H^2-m_h^2)c_\alpha s_\alpha+2m_{H^\pm}^2s_\beta c_\beta-m_{12}^2}{v^2s_\beta c_\beta},\\
\nonumber && \lambda_4 = \frac{(m_A^2-2m_{H^\pm}^2)s_\beta c_\beta+m_{12}^2}{v^2s_\beta c_\beta},\\
&& \lambda_5 = \frac{m_{12}^2 - m_A^2s_\beta c_\beta}{v^2s_\beta c_\beta}.
\label{eq:paratran}
\end{eqnarray}
One can see that we require an intolerably large $\ld_1\approx \tan^2\beta m_H^2/v^2 \gtrsim {\cal O}(10^4)$ in the large $\tan\beta$ region  if $m^2_{12}=0$. Thus, the soft $Z_2$ breaking term $m_{12}$ needs to be non-vanishing, and it is determined to be $m_{12}^2\approx m_{H}^2/ \tan\beta$. The mass splittings among the extra Higgs bosons are controlled by two parameters $\ld_{4,5}$:
\be\label{mass:sp}
m_H^2\approx m_A^2+\ld_5 v^2,\quad m_{H^+}^2\approx m_A^2+ \frac{1}{2}(\ld_5-\ld_4) v^2.
\ee
Immediately, we need $\ld_5\approx-\ld_4\sim {\cal O}(1)$ to get the favored mass pattern $m_A \ll m_H \simeq m_{H^\pm}$ by Electroweak precision test constraints. 
In addition, from Eq.~(\ref{eq:paratran}) we know that  in the large $\tan\beta$ limit we determine $\ld_2\approx m_h^2/v^2\approx 0.26$, just as in SM. 

In general, the Yukawa couplings of the five physical Higgs bosons, $h, H, A$ and $H^\pm$ in the 2HDM are given by
\begin{eqnarray}
\nonumber \mathcal L_{\mathrm{Yukawa}}^{\mathrm{2HDM}} &=&
-\frac{m_f}{v}\left(\xi_h^f\overline{f}hf +
\xi_H^f\overline{f}Hf - i\xi_A^f\overline{f}\gamma_5Af \right) \\
\nonumber &&-\left\{ \frac{\sqrt{2}V_{ud}}
{v}\overline{u}\left(m_{u}\xi_A^{u}P_L+m_{d}\xi_A^{d}P_R\right)H^{+}d  +
\frac{\sqrt{2}m_l}{v}\xi_A^l\overline{v}_LH^{+}l_R + \mathrm{H.C.}\right\},
\label{eq:L2hdm}
\end{eqnarray}
where $f$ runs over all of the quarks and charged leptons, and furthermore
$u$, $d$, and $l$ refer to the up-type quarks ($u$, $c$, $t$), down-type quarks ($d$, $s$, $b$),
and charged leptons ($e$, $\mu$, $\tau$), respectively. Specified to the L2HDM, we have
\begin{eqnarray}\label{effectiveC}
\nonumber &&\xi_h^u=\xi_h^{d}=\frac{\mathrm{cos}\,\alpha}{\mathrm{sin}\,\beta},\,\,
\xi_h^{l}=-\frac{\mathrm{sin}\,\alpha}{\mathrm{cos}\,\beta},\,\,\\
\nonumber &&\xi_H^u=\xi_H^{d}=\frac{\mathrm{sin}\,\alpha}{\mathrm{sin}\,\beta},\,\,
\xi_H^{l}=\frac{\mathrm{cos}\,\alpha}{\mathrm{cos}\,\beta},\,\,\\
&&\xi_A^{u}=-\xi_A^{d}=\mathrm{cot}\,\beta,\,\,
\xi_A^{l}=\mathrm{tan}\,\beta.
\label{eq:Ycoupling}
\end{eqnarray}
In any type of the 2HDM, the Higgs-to-gauge boson couplings read:
\begin{equation}
g_{hVV}=\mathrm{sin}(\beta-\alpha)g_{hVV}^{\mathrm{SM}},\,\,\,\,g_{HVV}=\mathrm{cos}(\beta-\alpha)g_{hVV}^{\mathrm{SM}},\,\,\,\,g_{AVV}=0,
\end{equation}
where $V$ refers to $Z$ and $W^\pm$ gauge bosons.
For very large value of tan$\,\beta$,
 we have $|\xi_H^{u,d}| \simeq |\xi_A^{u,d}| = \cot\beta$ and
 $|\xi_H^{l}| \simeq |\xi_A^{l}| = \mathrm{tan}\,\beta$, in short,
the quark Yukawa couplings of $H$ and $A$ are highly suppressed while
the lepton Yukawa couplings of $H$ and $A$ are highly enhanced.
This feature helps to shed a light on the muon {$g-2$} 
problem while evading various experimental constraints.

\section{Constraints on L2HDM parameters }
\label{sec:scan}

In this section we first describe all the relevant
theoretical and experimental constraints on the L2HDM parameter space.
Based on these constraints we present our results in 2-dimensional
profile likelihood maps. The 68\% (95\%) contours will be presented in 
dark (light) green in all the likelihood maps. 

\subsection{Enhanced $(g-2)_\mu $ with large $ \tan\beta$ and light $A$}

Recent progress in determining the muon anomalous magnetic moment $a_\mu = (g-2)_\mu/2$ establishes a 3$\sigma$ discrepancy:
\begin{equation}
\Delta a_\mu \equiv a_\mu^{\mathrm{EXP}} - a_\mu^{\mathrm{SM}} =  +262(85)\times 10^{-11},
\end{equation}
which is in a good agreement with various group's determinations \cite{broggio14}.
Such an excess can obviously be attributed to  a new physics contribution. In the framework of 2HDMs, the Barr-Zee 2-loop correction with a light $A$ and $\tau$ running in the loop \cite{bz} can generate a large positive $\Delta a_\mu$ due to an enhancement factor of $|\xi_A^l|^2 (m_\tau/m_\mu)^2$  in the large $\tan\beta$ limit. Let us note that the Barr-Zee diagram  with $H$ running in the loop gives  a negative contribution to $\Delta a_\mu$
and thus a heavier $H$ is preferred to enhance $\Delta a_\mu$. For more details, we refer the readers to Ref.~\cite{broggio14}.

We compute $(g-2)_\mu$ by using package \texttt{2HDMC}~\cite{2hdmc}.\footnote{Alternative option is 
the public \texttt{Mathematica} code~\cite{Queiroz:2014zfa}.}

\subsection{Theoretical constraints}

There are several theoretical constraints; the perturbativity, vacuum stability and unitarity bounds to be considered.
All of them are implemented at the weak scale. In particular, the first imposes the highest mass scale for the Higgs states.

\begin{itemize}

\item For the perturbativity, we put the bound: $|\lambda_i|<4\pi$ for $i$=1,...,5.

An immediate consequence of this bound can be obtained from Eq.~(\ref{mass:sp}):
\begin{equation}
m_{H,H^\pm} ^2<4\pi v^2+m_A^2,
\label{H:upper}
\end{equation}
saturated for $\ld_{5}\simeq-\ld_4=4\pi$.
Assuming a small contribution from $m_A$, it gives the upper bound  $m_{H^+}\sim m_H\lesssim900\,$GeV.
Note that with the large tan$\,\beta$ approximation, $\lambda_1$ becomes an independent parameter
and its magnitude is in principle allowed to run within $4\pi$ by perturbativity.

\item Vacuum stability demands
\begin{equation}
\lambda_{1,2}>0,\,\,\lambda_3+\sqrt{\lambda_1\lambda_2}>0,\,\,|\lambda_5|< \lambda_3+\lambda_4+\sqrt{\lambda_1\lambda_2}.
\label{eq:vaccstab}
\end{equation}
The last condition  can be rewritten as $\ld_{3}+\ld_4-\ld_5>-\sqrt{\ld_1\ld_2} $ for $m_H>m_A$.
One of the key features in our discussion is that the couplings and thus the upper limits on the heavy Higgs masses show quite different behaviors in the right-sign (SM) and wrong-sign limit of the normalized Yukawa coupling $\xi^l_h$ \cite{ferreira14}. Using a trigonometric identity, $\xi^l_h$ can be expressed by
\begin{equation} \label{xitau}
\xi^l_h = -{s_\alpha\over c_\beta} \equiv s_{\beta-\alpha}-t_\beta c_{\beta-\alpha}.
\end{equation}
As found at the LHC, the 125 GeV Higgs boson $h$ is very much SM-like requiring, in particular,  $|s_{\beta-\alpha}| \simeq 1$ and $|\xi^\tau_h|\approx 1$.  Notice that this can be reached
in the SM limit $t_\beta c_{\beta-\alpha} \approx 0$ (leading to the right-sign lepton coupling 
$\xi^l_h \approx +1$),
or in the large $\tan \beta$ limit with  $t_\beta c_{\beta-\alpha}\approx 2$ (leading to the wrong-sign couplig $\xi^l_h \approx -1$). Using the relation (\ref{xitau}), one finds
\begin{equation}
 \lambda_3 + \lambda_4 -\lambda_5 =
 { 2 m_A^2 + \xi_h^l s_{\beta-\alpha} m_h^2 - (s_{\beta-\alpha}^2 + \xi_h^l s_{\beta-\alpha}) m_H^2
 \over v^2} +{\cal O}({1\over t^2_\beta})
 \label{lam345}
 \end{equation}
 in the large $\tan\beta$ limit.
Now, in the right-sign limit ($\xi_h^l s_{\beta-\alpha} \to +1$), we have
\begin{equation}
2 {m_H^2 \over v^2} < \sqrt{0.26 \times 4\pi} + {2 m_A^2 + m_h^2 \over v^2}
\end{equation}
which puts a bound $m_H < 250$ GeV  for $m_A=0$, which is consistent with~\cite{broggio14}.
On the other hand, in the wrong-sign
limit ($\xi_h^l s_{\beta-\alpha} \to -1$), $m_H$ can be arbitrarily large allowing a fine-tunnig 
$s_{\beta-\alpha}^2+\xi^l_h s_{\beta-\alpha} \approx 0$. 
These properties will be clearly shown in our Figs.~2 and 3.

\item Tree-level unitarity for the scattering of Higgs bosons and the longitudinal parts of the EW gauge bosons.

The numerical evaluation of the necessary and sufficient conditions for the tree-level unitarity in the general 
2HDM has been encoded by the open-source program \texttt{2HDMC}~\cite{2hdmc}.
We deal with this constraints relying on it.
Here, we point out that this constraint is rather loose in the following reason.
In the limit of large tan$\,\beta$, the  parameter $\lambda_1$ decouples from the 
other  parameters $\lambda_{2,3,4,5}$, and is allowed to run between 0 and $4\pi$ independently.
Therefore, one can always track down a value of $\lambda_1$ 
to meet the requirement of the tree-level unitarity without 
affecting any  other physical observables significantly.
\end{itemize}

\subsection{Electroweak precision test}

Electroweak precision test (EWPT), commonly referred to as the $\rho$ parameter bound, is taken into account by calculating the oblique parameters, $S,T$ and $U$ in the \texttt{2HDMC} code. As we are interested in a splitting spectrum of $A$ and $H,\,H^\pm$, the custodial symmetry is potentially violated significantly. However, as analyzed in detail in Ref.~\cite{broggio14}, 
taking the SM limit $s_{\beta-\alpha}\to 1$, the custodial symmetry can be restored 
if $m_{H^\pm} \approx m_H (m_A) $ for arbitrary value of $m_A (m_H)$~\cite{gerard07}.
In our scan study, we reproduce the previous results as clearly demonstrated in Fig.~\ref{degeneracy}. 
Let us remark that we have updated the central values, error bars, and correction matrix adopted in Ref.~\cite{broggio14}, using the new PDG data~\cite{Agashe:2014kda}.

\subsection{ Light $A$ and Higgs exotic decay\label{sec:EXOdecay}}

As we are interested in the case of a light CP-odd scalar $A$, the SM Higgs boson
can have an exotic decay of (i) $h\to AA$ for $m_A<m_h/2$, or (ii) $h\to A A^* (\tau^+ \tau^-)$
for  $m_A>m_h/2$~\footnote{In type-I and type-II 2HDM, Ref.~\cite{Bernon:2014nxa} studied the possibility of two-body decay mode $h\ra AA$ while the three-body decay mode was ignored.}. At the moment, the current LHC data on the SM Higgs boson put a strong constraint on the $hAA$ coupling $\lambda_{hAA}$ and $m_A$.  On the other hand, it will be an interesting channel to test the hypothesis of the L2HDM explaining the muon {$g-2$} at the next runs of the LHC.
The partial decay widths of these processes are
\begin{eqnarray}\label{decay:hAA}
&&\mbox{(i)} ~~ \Gamma(h\ra AA) = \f{1}{32\pi}\f{\lambda^2_{hAA}}{m_h}
\sqrt{1-4 m_A^2/m_h^2},  \\
&&\mbox{(ii)} ~~ \Gamma(h\ra A\tau \tau) \approx \f{1}{128\pi^3}
\f{\lambda^2_{hAA} m^2_\tau}{m_h v^2} \tan^2\beta\, G(m_A^2/m_h^2),\\
&& ~~\mbox{where}~
G(x) \equiv (x-1)\L 2-\f{1}{2}\log x \R+\f{1-5x}{\sqrt{4x-1}}\L \arctan \f{2x-1}{\sqrt{4x-1}}-\arctan \f{1}{\sqrt{4x-1}} \R. \nonumber
\end{eqnarray}
The function $G(x)$ is a very fast monotonically decreasing function with respect to $x$. For instance,
we have $G(0.3)\approx0.28$ to be compared with $G(0.5)\approx0.0048$.

Generically, $\lambda_{hAA}$ is expected to be around the weak scale hence leading to a large decay width at the GeV scale,
which is readily excluded. 
To avoid this situation, one may require $m_A>m_h/2$ or arrange a mild cancelation 
to get  sufficiently small $\lambda_{hAA}$. 
Interestingly, one can find
\begin{equation}\label{haa:1}
\lambda_{hAA} \approx - (\ld_3+\ld_4-\ld_5) v,
\end{equation}
where $\lambda_3 + \lambda_4-\lambda_5$ is given in Eq.~(\ref{lam345}). This relation says that
there could be a cancellation among three contributions from $m_A, m_h$ and $m_H$. In particular, for $m_H \gg m_{h,A}$ of our interest, the cancellation is obtained only in the wrong-sign limit with $\xi_h^l \lesssim -1$.
This can be explicitly seen by taking $\lambda_{hAA}$ as a free parameter (traded with $\lambda_1$) and expressing  the normalized tau (lepton) coupling as
\begin{equation} \label{xi_tau}
 \xi_h^l s_{\beta-\alpha} \approx  -{ s^2_{\beta-\alpha} m_H^2 -2 m_A^2 - v \lambda_{hAA}/s_{\beta-\alpha}  \over m_H^2 -m_h^2}.
\end{equation}
 {In the limit of $m_H \gg m_{A}$ and $\lambda_{hAA}\to 0$, it can be further approximated as $-m_H^2/(m_H^2-m_h^2)\lesssim -1$, and  thus we have $ \xi_h^l \lesssim -1$.~\footnote{ {The case with $s_{\beta-\alpha}\approx-1$ (or equivalently $\cos\alpha\approx-1$), i.e., the reversed couplings of other SM couplings, is completely excluded from our numerical results. So, we have $s_{\beta-\alpha}\approx+1$ in this paper.}}} We demonstrate this behavior in the right panel of Fig.~\ref{haa}.
 
The presence of a light $A$ may leave hints at Higgs exotic decay through the channel $h\ra AA(A^*)\ra$4$\tau$. 
The upper bound of the exotic branching ratio of the Higgs decay is known to be 60~\%, 
however, a mildly more stringent bound on the $h\ra AA$ mode 
using multilepton searches by CMS~\cite{CMS:AA} can be set:
Br$(h\ra AA\ra 4\tau)\lesssim 20\%$ almost independent on $m_A$~\cite{Curtin:2013fra}.
In this paper we impose a conservative cut Br$(h\ra AA(A^*))\lesssim 40\%$.

\subsection{Collider and other constraints}

\begin{itemize}

\item Collider searches on the SM and exotic Higgs bosons

For various Higgs constraints from LEP, Tevatron and LHC, we rely on the package \texttt{HiggsBounds-4.2.0}~\cite{Bechtle:2013wla}
incorporating the most updated data on $BR(h\to\tau\tau)$. 
We notice that the DELPHI search~\cite{delphi04} on the process
\begin{align}\label{LEP:A}
e^+e^-\ra Z^* \ra A H \ra 4\tau,
\end{align}
is sensitive to our model. The Fig. 15 in the Ref.~\cite{delphi04} shows 
the region $m_A+m_H\lesssim 185\gev$ is excluded at 95\% confidence level.

Specific to our study, the 125 GeV Higgs decay $h\ra \tau^+\tau^-$ is of particular concern as it can deviate
significantly from $\pm1$ as indicated in Eq.~(\ref{xi_tau}).
We use the new data from CMS~\cite{Khachatryan:2014jba} and ATLAS~\cite{Aad:2015vsa},
weighted by their statistic error bars:
\begin{equation}
\mu_{\tau\tau} = \begin{cases} 1.43\pm0.40 ~~\mbox{ATLAS} \cr
                                          0.91\pm0.28 ~~\mbox{CMS} \end{cases}.
\end{equation}

\item $B_s \to \mu^+ \mu^-$

The light $A$ contribution to the decay $B_s \to \mu^+ \mu^-$ becomes sizable if $m_A \lesssim 10$ GeV.  In our analysis, we do not include this constraint as it is irrelevant for $m_A > 15$ GeV.  More details can be found in Refs.~\cite{wang14,abe15}.

\item $\tau$ decays and lepton universality

In the limit of light $H^\pm$ and large $\tan\beta$, the charged Higgs boson can generate 
significant corrections to $\tau$ decays at tree and 1-loop level~\cite{maria04}. Recent study \cite{abe15} 
attempted to put a stringent bound on the charged Higgs contributions from 
the lepton universality bounds obtained by the HFAG collaboration \cite{hfag14}. 
Given the precision at the level of  0.1 \%, the HFAG data turned out to provide most stringent bound on the L2HDM parameter space in favor of the muon $g-2$.  Thus, it needs to be considered more seriously. For this, we improve the previous analysis treating the HFAG data in a proper way.

From the measurements of the pure leptonic processes, $\tau \to \mu \nu \nu$, $\tau \to e \nu \nu$
 and $\mu \to e \nu\nu$, HFAG obtained the constraints on the three coupling ratios, 
 $(g_\tau/g_\mu) = \sqrt{ \Gamma( \tau \to e  \nu\nu)/\Gamma(\mu \to e \nu\nu)}$, etc.
Defining $\delta_{l l'}\equiv (g_l/g_{l'})-1$, let us rewrite the data:
\begin{equation} \label{LUdata1}
\delta^l_{\tau\mu} =  0.0011 \pm 0.0015,\quad
\delta^l_{\tau e} = 0.0029 \pm 0.0015, \quad
\delta^l_{\mu e} = 0.0018 \pm 0.0014
\end{equation}
In addition, combing the semi-hadronic processes $\pi/K \to \mu \nu$, HFAG  also provided 
the averaged constraint on $(g_\tau/g_\mu)$ which is translated into
\begin{equation} \label{LUdata2}
\delta^{l+\pi+K}_{\tau\mu} = 0.0001 \pm 0.0014.
\end{equation}
We will impose the above lepton universality constraints in our parmeter space. 

Now, it is  important to notice that only two ratios out of three leptonic measurements are independent
and thus they are strongly correlated as represented by the correlation coefficients \cite{hfag14}. Therefore, one combination of the three data has to be projected out. One can  easily check that the direction $\delta^l_{\tau\mu}-\delta^l_{\tau e} + \delta^l_{\mu e}$ has the zero best-fit value and the zero eigenvalue of the covariance matrix, and thus corresponds to the unphysical direction.
Furthermore, two orthogonal directions $\delta^l_{\tau\mu} + \delta^l_{\tau e}$ and $-\delta^l_{\tau\mu}+\delta^l_{\tau e} + 2 \delta^l_{\mu e}$ are found to be uncorrelated in a good approximation.  In the L2HDM, the deviations from the SM prediction $\delta_{l l'}$ are calculated to be
\begin{equation} \label{deltas}
\delta^l_{\tau\mu} = \delta_{loop},\quad
\delta^l_{\tau e} = \delta_{tree} + \delta_{loop},\quad
\delta^l_{\mu e} = \delta_{tree}, \quad
\delta^{l+\pi+K}_{\tau \mu} = \delta_{loop}.
\end{equation}
Here $\delta_{tree}$ and $\delta_{loop}$ are given by \cite{maria04}:
\begin{eqnarray} \label{delta_tree_loop}
\delta_{tree} &=& {m_\tau^2 m_\mu^2 \over 8 m^4_{H^\pm}} \tan^4 \beta
- {m_\mu^2 \over m^2_{H^\pm}} t^2_\beta {g(m_\mu^2/m^2_\tau) \over f(m_\mu^2/m_\tau^2)}, \\
\delta_{loop} &=& {G_F m_\tau^2 \over 8 \sqrt{2} \pi^2} t^2_\beta 
\left[1 + {1\over4} \left( H(x_A) + s^2_{\beta-\alpha} H(x_H) + c^2_{\beta-\alpha} H(x_h)\right)
\right], \nonumber
\end{eqnarray}
where $f(x)\equiv 1-8x+8x^3-x^4-12x^2 \ln(x)$, $g(x)\equiv 1+9x-9x^2-x^3+6x(1+x)\ln(x)$, 
$H(x) \equiv \ln(x) (1+x)/(1-x)$, and $x_\phi=m_\phi^2/m_{H^{\pm}}^2$.
From Eqs.~(\ref{LUdata1},\ref{LUdata2}) and (\ref{deltas}), one obtains the following three independent bounds: 
\begin{eqnarray} \label{LUconstraints}
 {1\over\sqrt{2}} \delta_{tree} +\sqrt{2} \delta_{loop} &=& 0.0028 \pm 0.0019,\quad \nonumber\\ 
{\sqrt{3\over2} } \delta_{tree} &=& 0.0022\pm 0.0017,\quad \\
\delta_{loop} &=&0.0001 \pm 0.0014 \nonumber.
\end{eqnarray}

\begin{figure}[h]
\includegraphics[width=3.5in]{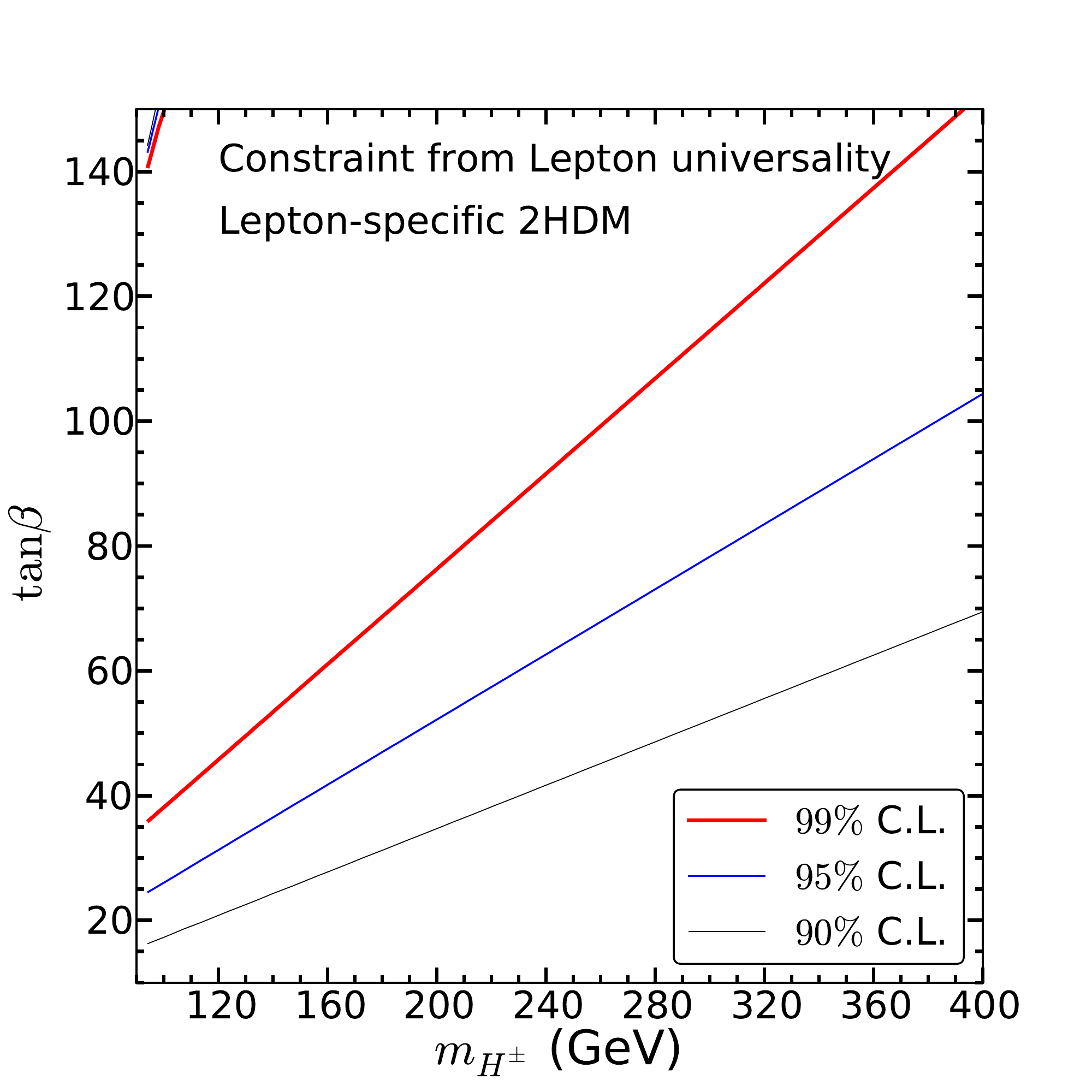}
\caption{The contours of lepton universality likelihood profiled on ($m_{H^\pm}$, $\tan\beta$) plane. 
The red, blue, and black lines are corresponding to $99\%$, $95\%$, and $90\%$ confidence limit, respectively.}
\label{fig:LU}
\end{figure}

Based on the constraints Eq.~(\ref{LUconstraints}) on the two fundamental free parameters $\delta_{tree}$ and $\delta_{loop}$, 
we can draw the the lepton universality likelihood contours, where 
we found the minimum value $\chi^2_{\min}=0.229$. 
In Fig.~\ref{fig:LU}, 
we present profile likelihood contours on the $m_{H^\pm}$-$\tan\beta$ plane
the red, blue, and black lines are 
corresponding to $99\%$, $95\%$, and $90\%$ confidence level, respectively.  
Note that the $\delta_{loop}$ is always negative in the region of our interest listed in Table~\ref{tab:inputs}.
On the other hand, $\delta_{tree}$ depends only on the parameter $\tan\beta/m_{H^\pm}$ and 
negative in most of the region but can be also positive. In a fine-tuned region located $\tan\beta/m_{H^\pm} \sim 1$~GeV$^{-1}$
as we can see in the large $\tan\beta$ and small $m_{H^\pm}$ corner in Fig.~\ref{fig:LU}, where the positive $\delta_{tree}$ and the negative $\delta_{loop}$ cancel.

We also found that lepton universality likelihood is practically not sensitive to the heavy neutral Higgs mass $m_{H}$  
and $\cos(\beta-\alpha)$ in our region of interest. Hence, we show the lepton universality contours only on the $m_{H^\pm}$-$\tan\beta$ plane (Fig.~\ref{fig:LU})
and  on the $m_{A}$-$\tan\beta$ plane (Fig.~\ref{distributions} left panel).

\end{itemize}

Let us finally remark that  we use Gaussian distribution or hard cut for the likelihood 
functions to impose the experimental constraints. 
When the central values, experimental errors and/or theoretical errors are available, 
Gaussian likelihood is used. Otherwise the hard cut is adopted such as the Higgs limits implemented
in \texttt{HiggsBounds}.

\begin{table}[h!]
\begin{center}
\begin{tabular}{|c|c|}
\hline
2HDM parameter & Range \\
\hline
Scalar Higgs mass $(\gev)$&  $125<m_H<{400}$ \\
Pseudoscalar Higgs mass $(\gev)$& $10<m_A<400$ \\
Charged Higgs mass $(\gev)$& $94<m_{H^\pm}<400$ \\
$ c_{\beta-\alpha}$& {$0.0< c_{\beta-\alpha}<0.1$} \\
$ \tan\beta$ & $10< \tan\beta <150$ \\
$\lambda_1$ & $0.0<\lambda_1<4\pi$ \\
\hline
\end{tabular}
\caption{The scan ranges of the input parameters over which we perform the scan of L2HDM. 
Note that we adopt the convention in  \texttt{2HDMC}; 
$-\pi/2<\alpha-\beta<\pi/2$ and $0<\beta<\pi/2$, and use
 the parameter $\lambda_1$ as an input parameter instead of
$m^2_{12}$ in order to make the scan more efficient.
}
\label{tab:inputs}
\end{center}
\end{table}

\subsection{Results}

Our input parameters and the scan ranges of them are summarized in Table~\ref{tab:inputs}.  
Some comments are in order.
(i) We focus on the case that the SM-like Higgs boson $h$ is the lighter CP-even Higgs boson 
with mass $125\gev$~\cite{Aad:2015zhl}. 
\footnote{We have checked the case that the SM-like Higgs is the heavier CP-even Higgs. 
We found that the allowed region is rather restricted at $m_h\simeq m_H\simeq 125\gev$, which 
is the similar solution to the subset of region (B).  
}
(ii) We require  $\cos(\alpha-\beta)\leq 0.1$,
which guarantees that $h$ couples to quarks and vector bosons without
appreciable deviation from the SM predictions.
The updated LHC results can be found in Ref.~\cite{atlas14010}.
(iii) The upper bound on $m_{H, H^\pm} <400$ GeV is put by hand
since we are interested in the relatively light region testable at the LHC near future.
In principle, they can be as heavy as about 900 GeV according to the perturbativity constraints.
(iv) We restrict ourselves to $ \tan\beta\leq 150$.

\begin{figure}[b]
\includegraphics[width=3.1in]{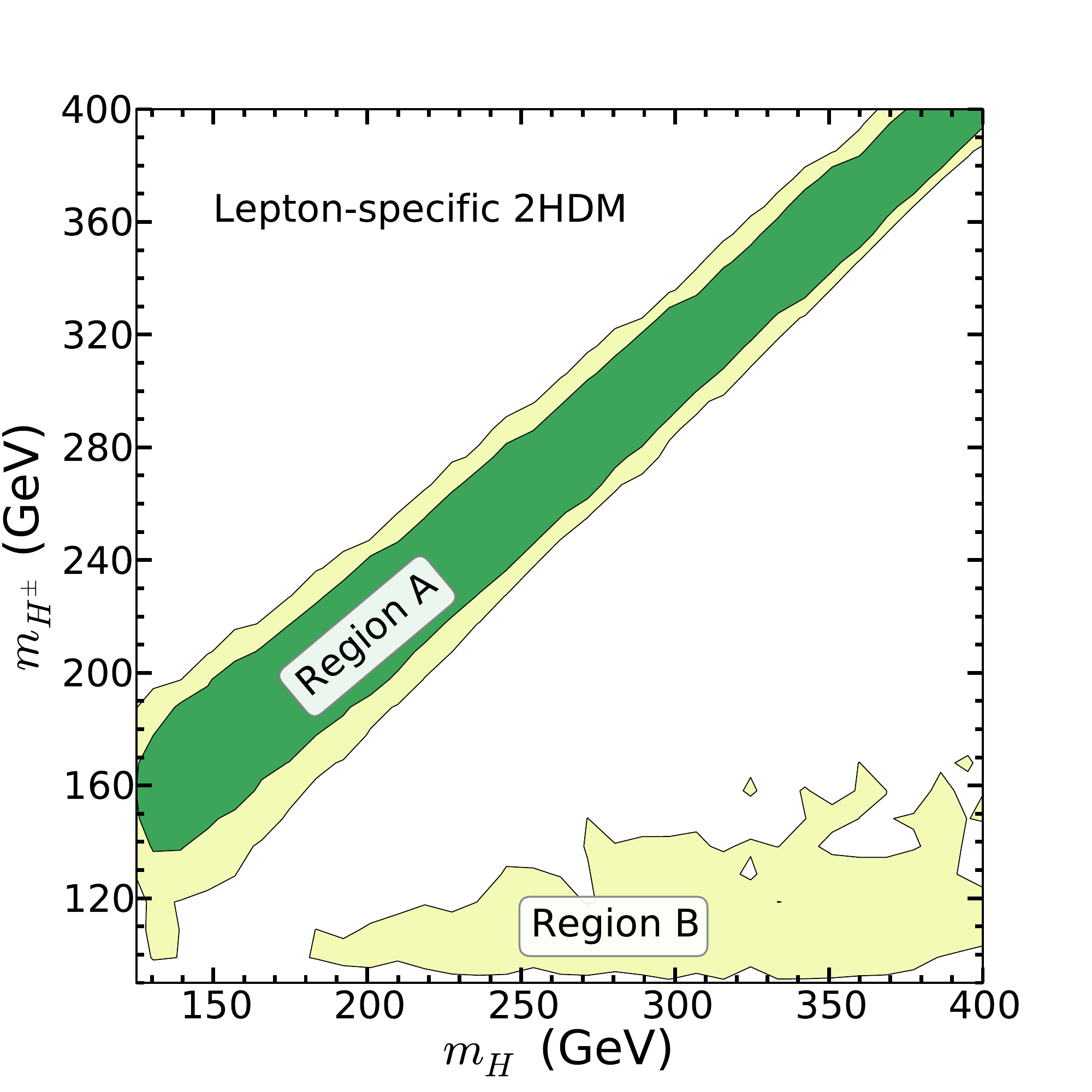}
\includegraphics[width=3.1in]{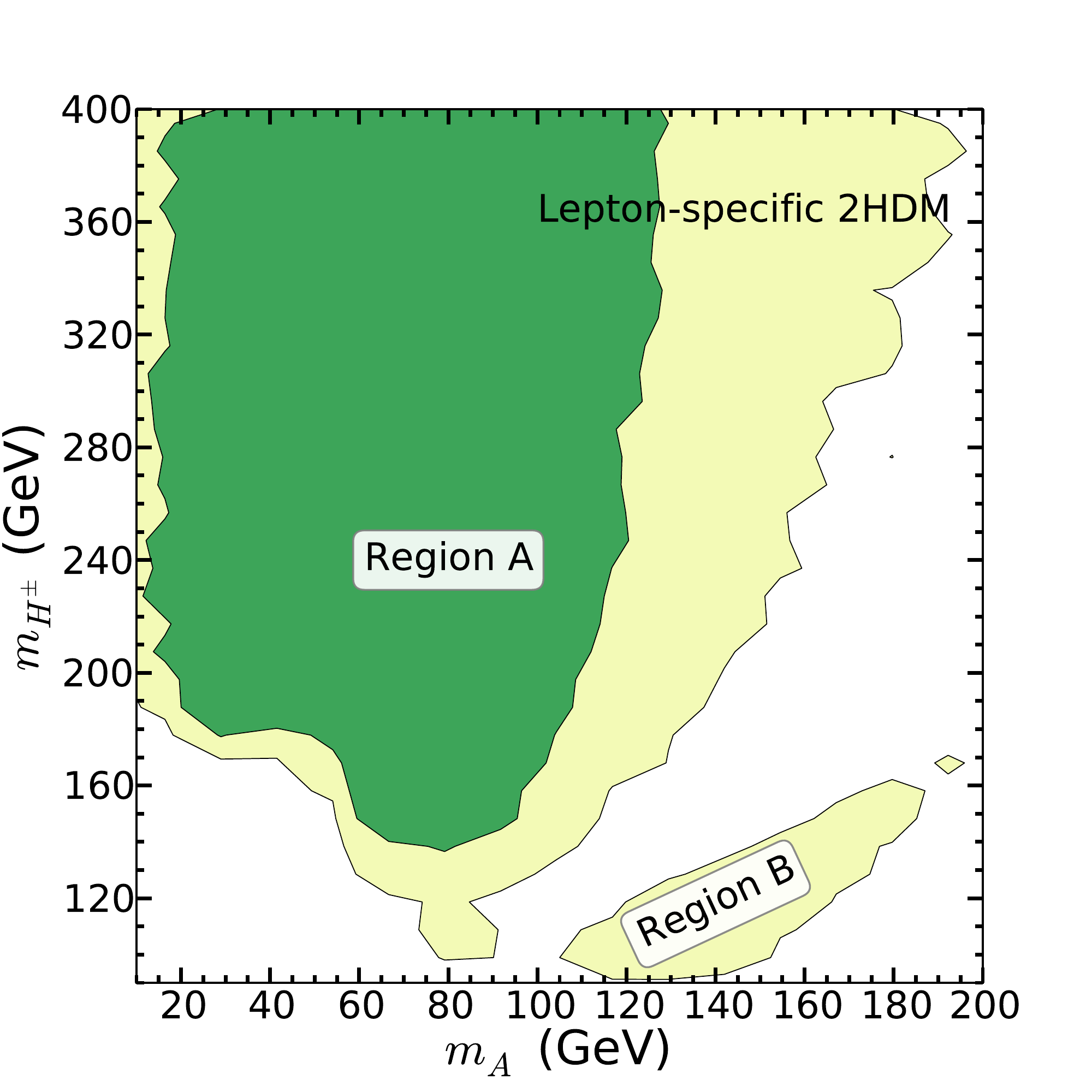}
\caption{Features of the Higgs spectrum with a light $A$ facing EWPT.
The inner green (outer light green) contours are $68\% \, (95\%)$
confidence region.
Distribution on the $m_H-m_{H^\pm}$ plane (left) and the $m_A-m_{H^\pm}$ plane (right). }
\label{degeneracy}
\end{figure}

\begin{figure}[htb]
\includegraphics[width=3.1in]{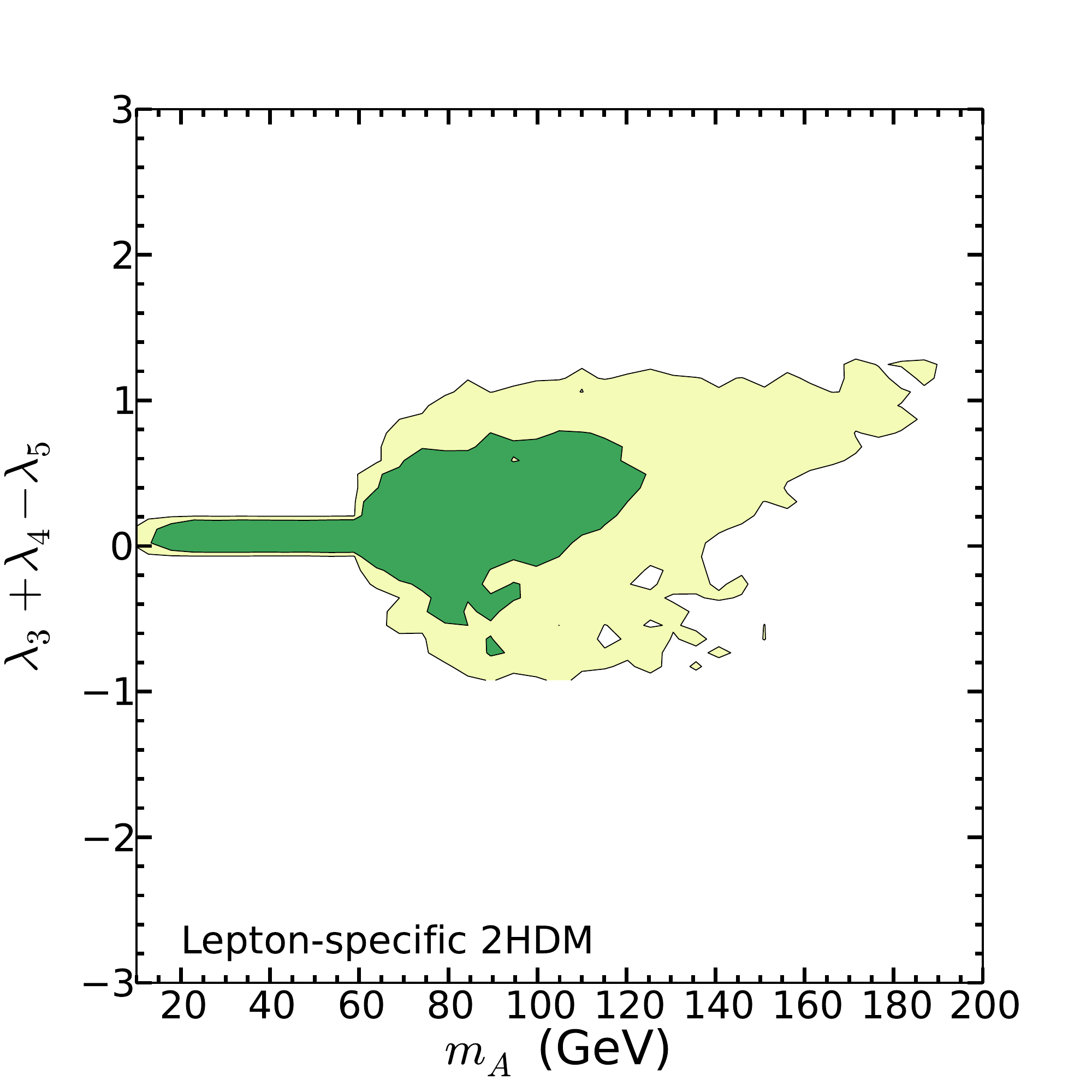}
\includegraphics[width=3.1in]{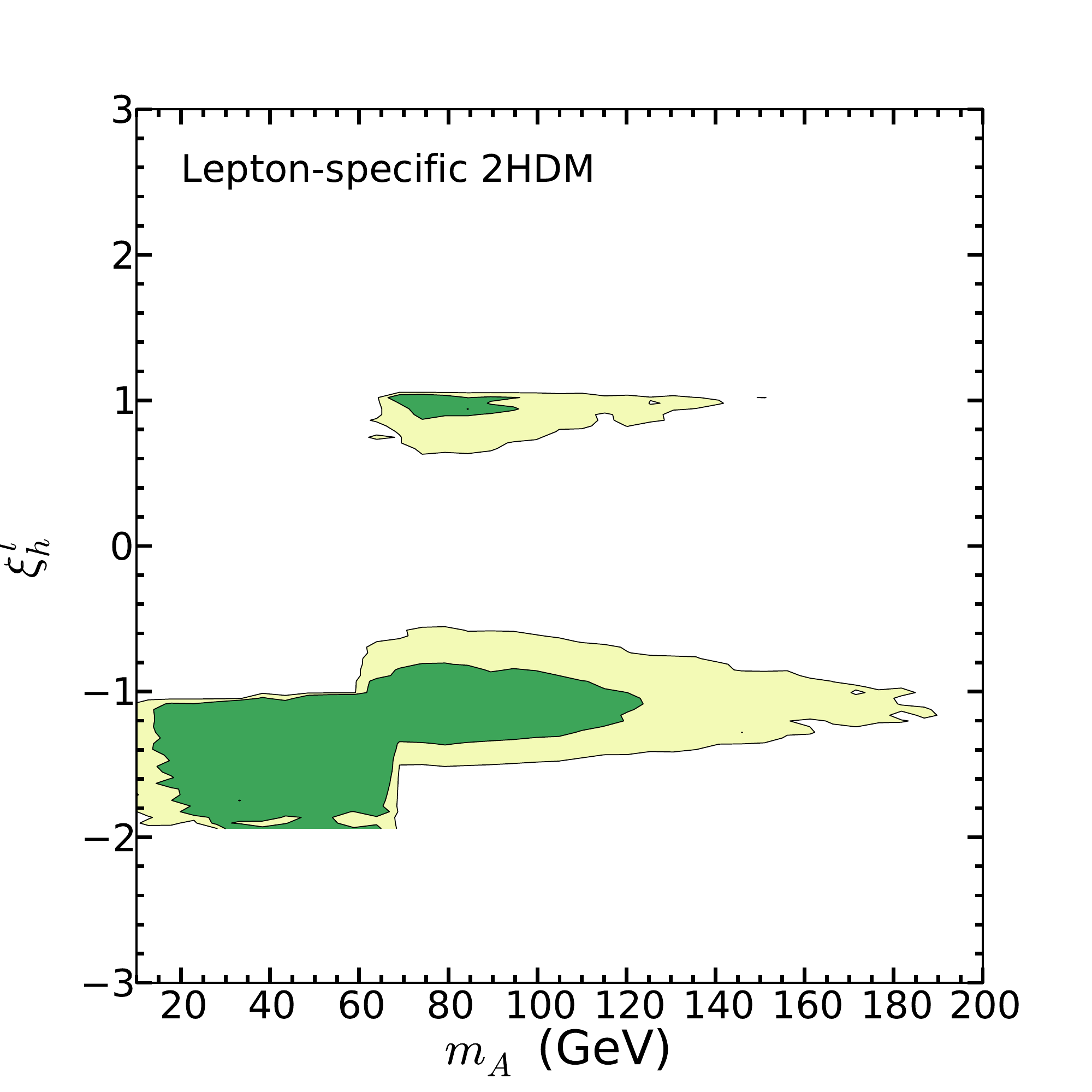}
\caption{
The 2-dimensional profile likelihood. 
The inner green (outer light green) contours are $68\% \, (95\%)$
confidence region.
Left panel: the coupling $\mu_{hAA}$ (in unit of $v$) versus $m_A$.
Right panel: the reduced coupling of leptons $\xi_h^l$ versus $m_A$.
}\label{haa}
\end{figure}

\begin{figure}[htb]
\includegraphics[width=3.1in]{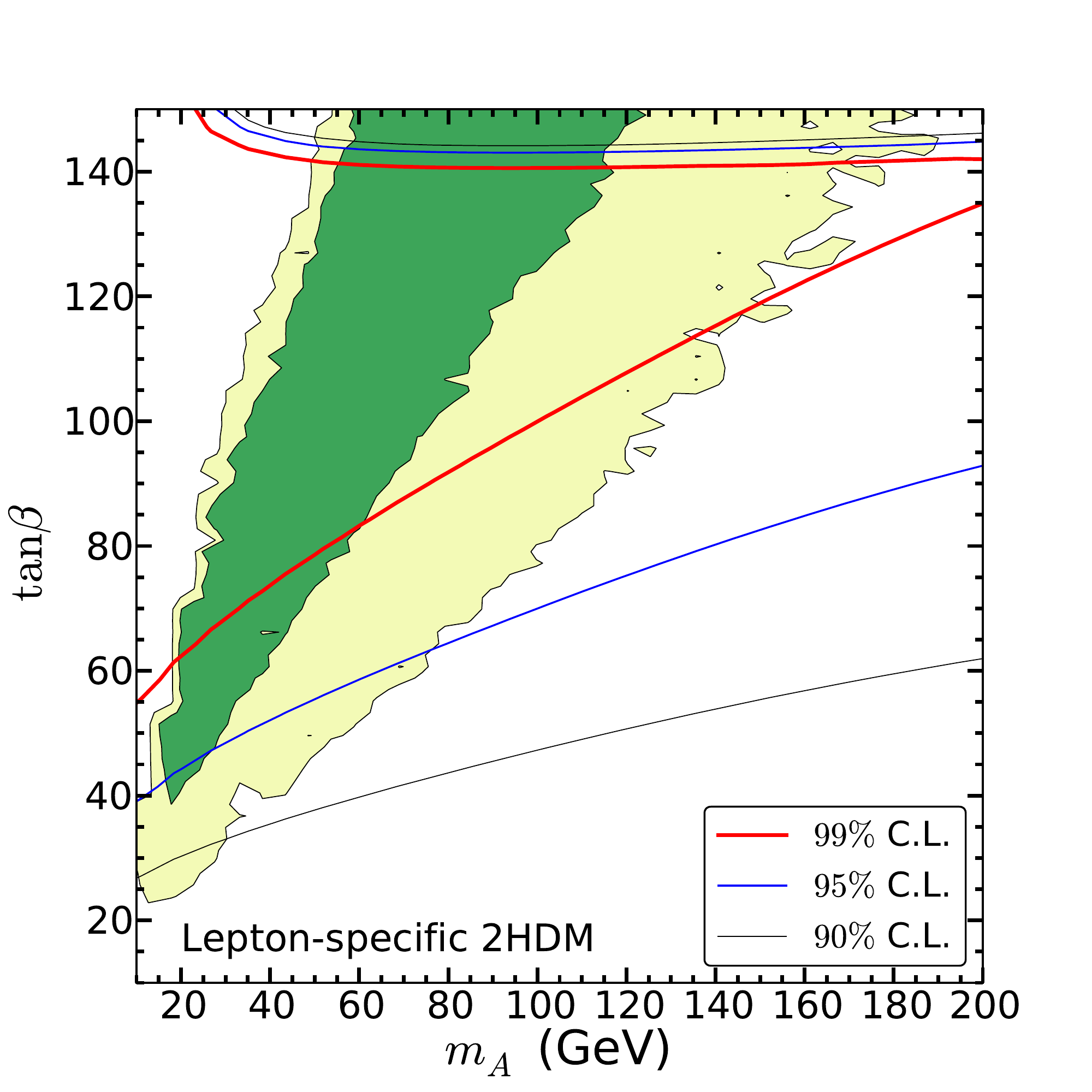}
\includegraphics[width=3.1in]{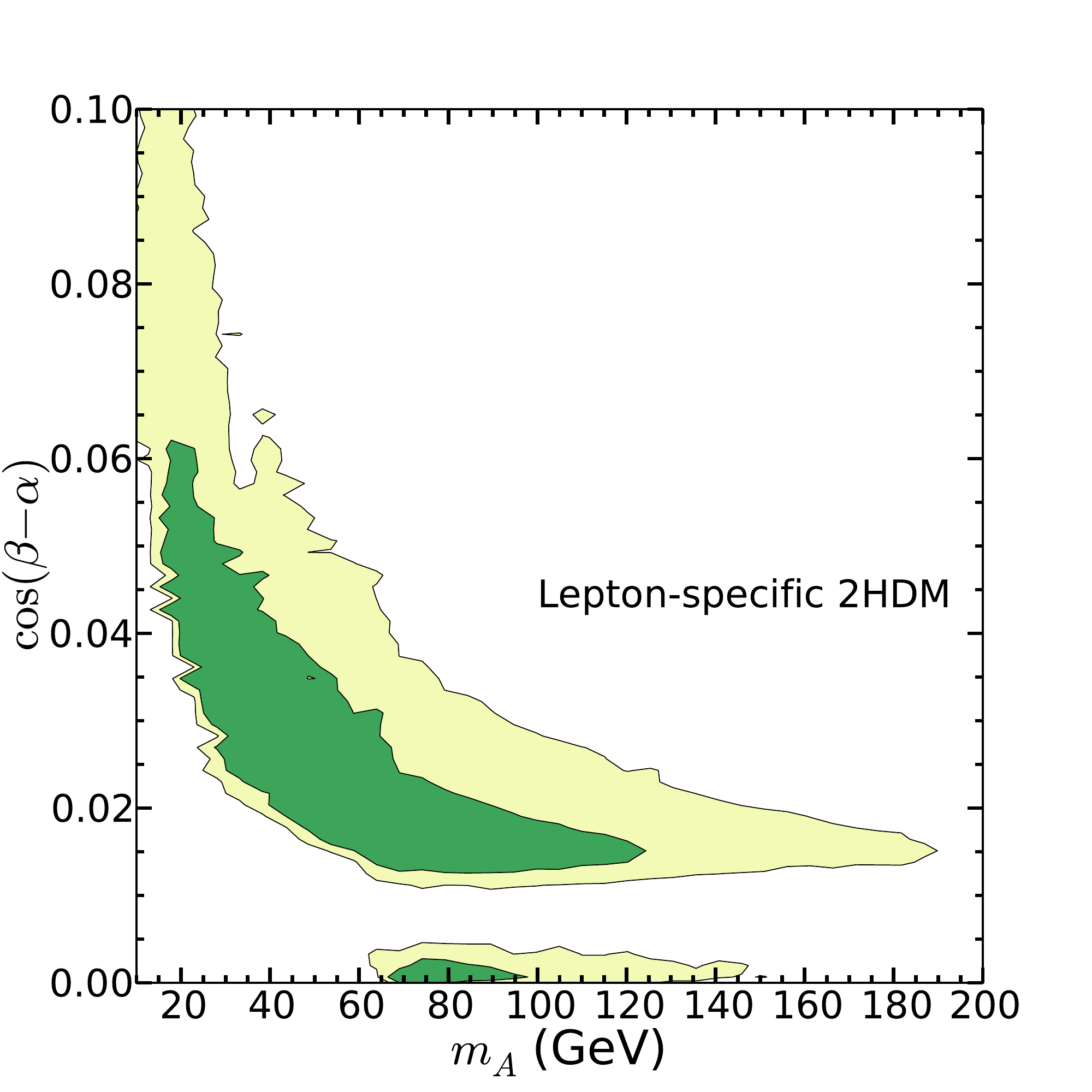}
\caption{Left: Distribution on the $m_A- \tan\beta$ plane (left), and the $m_A - \cos(\alpha-\beta)$
plane (right). 
The contours of lepton universality likelihood are also presented in $99\%$ (red), 
$95\%$ (blue), and $90\%$ (black) confidence limit. }
\label{distributions}
\end{figure}

\begin{figure}[htb]
\includegraphics[width=3.1in]{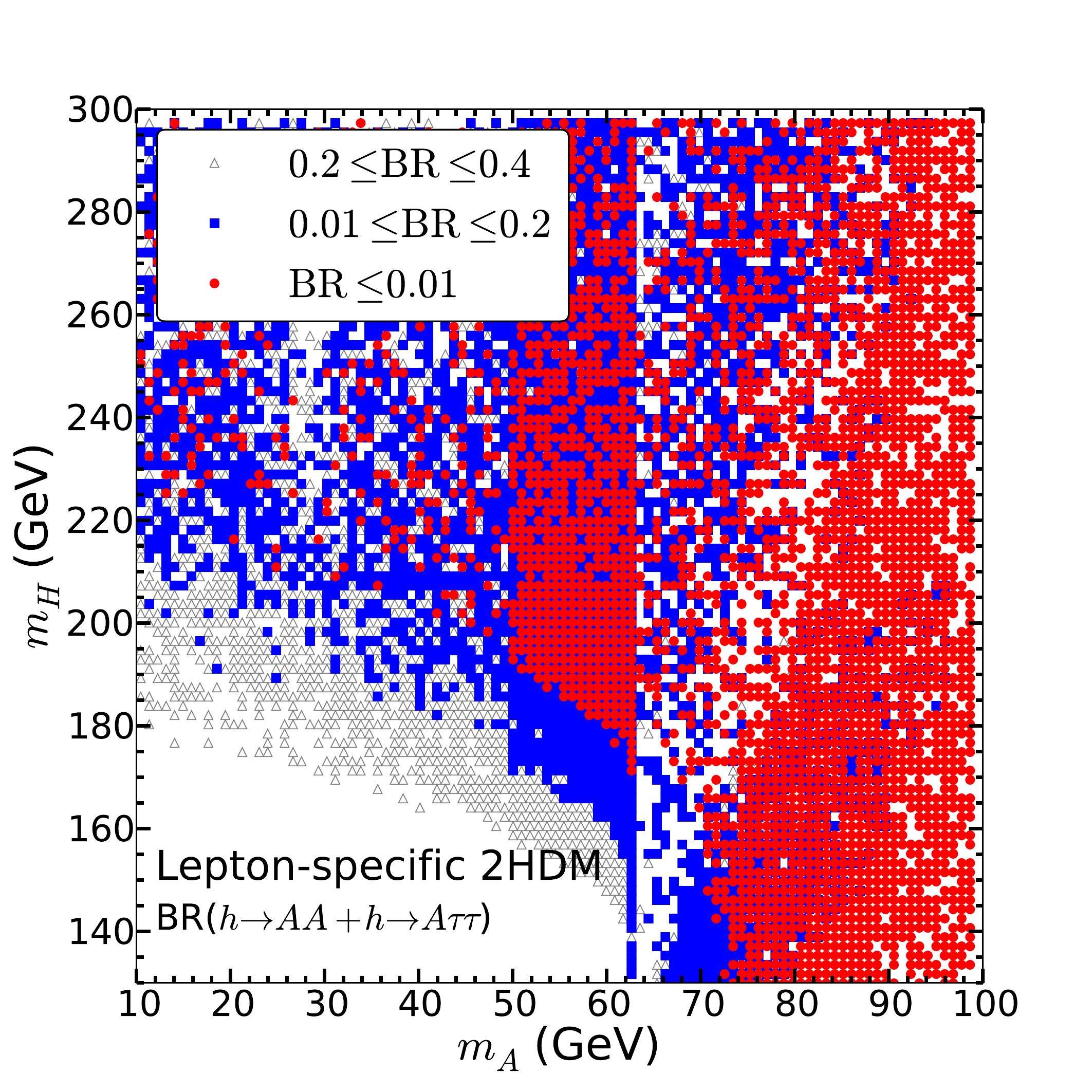}
\caption{Plots of the SM-like Higgs exotic decay Br$(h\ra AA)$ (for $m_A\lesssim m_h/2$) and Br$(h\ra A\tau^+\tau^-)$  (for $ m_h/2\lesssim m_A\lesssim m_h$.).
All the scatter points satisfy the constraints described in the text in $2\sigma$.  
}\label{EXOdecay}
\end{figure}

We show the scan results in several 2 dimensional profile likelihood maps 
from Fig.~\ref{degeneracy} to Fig.~\ref{distributions}. 
The inner green (outer light green) contours are $68\% \, (95\%)$
confidence region.
The points are summarized in the following:

\begin{itemize}

\item  The left panel of Fig.~\ref{degeneracy} shows two separated allowed regions.
The majority is crowding around the line $m_{H}= m_{H^+}$, which is in well accordance with the EWPT via accidental degeneracy between $H$ and $H^\pm$. Note that there is a lower bond on $m_H\sim m_{H^+}$, about 130 GeV. 
The minority is on the small island with quite light ${H^\pm}$ near  $m_{H^\pm} \sim 100$ GeV, just in the vicinity of the LEP bound on charged particles. With the help of the right panel of Fig.~\ref{degeneracy}, one finds a mild degeneracy between $A$ and $H^\pm$ with $m_A \approx 100-180$ GeV and  $m_{H^\pm}\lesssim 160$ GeV. For $m_A >100$ GeV,  $\tan\beta$ needs to be larger than about  70, see Fig.~\ref{distributions}. We call the former region as Region A and the latter as Region B.
Note that the fragmentation of the plots, particularly in the region B 
of the left panel of Fig.~\ref{degeneracy}, is due to a coarse-tuning likelihood.
As we will see in the next section, Region B is already excluded by the current LHC 8TeV data.

\item
The left panel of Fig.~\ref{haa} shows the relation between $\lambda_{hAA}$ and $m_A$.
We see only $|\lambda_{hAA}|\sim 0$ is allowed for $m_A\lesssim 60\gev$, 
while larger $|\lambda_{hAA}|$ is allowed for 
$m_A\gtrsim 60\gev$.
The right panel of Fig.~\ref{haa} shows the relation between $\xi^{\tau}_h$ vs. $m_A$.
In the region $m_A\lesssim 70\gev$, 
only the wrong-sign region ($\xi_h^l<0$) is allowed. It is consistent 
with suppressed $\lambda_{hAA}$ seen in the left panel as discussed in Eq.~(\ref{xi_tau}). 
For heavier $A$, there appears the right-sign region.

\item Remarkably, the $m_A\lesssim 60$ GeV region tends to show an enhancement in Br$(h\ra \tau\tau)$, up to a factor $|\xi_{h}^l|^2 \sim 4$. While above it both (mild) enhancement and suppression are possible. Further precise measurement of  Br$(h\ra \tau\tau)$ helps to shrink the allowed parameter regions.

\item In the left panel of Fig.~\ref{distributions}, 
The contours of lepton universality likelihood are also presented in $99\%$ (red), 
$95\%$ (blue), and $90\%$ (black) confidence limit. 
The region with $\tan\beta < 140$ with small $m_A$ allowed by other constraints are very
constrained by lepton universality. 
However, the region located at the large $\tan\beta > 140$ are 
always allowed by the fine-tuning cancellation between $\delta_{tree}$ and $\delta_{loop}$
by selecting an appropriate $m_{H^\pm}$. 
The lower $\tan\beta$ region allowed at 95\% appears to be a consistent combination 
of the same 95\% contour lines with different values of $m_{H^\pm}$ in~\cite{abe15}.

\item A light $A$ with $m_A \sim20-63$ GeV is of our particular interest.\footnote{Remark again this region is further reduced by considering the tau decay and lepton universality data~\cite{abe15}.}
In this region, the wrong-sign limit
($\xi_h^l \sim -1$) has to be realized and thus the lower bound on $\tan\beta$ is correlated
with the upper bound on $\cos(\alpha-\beta)$, 
which can be seen from the right panel of Fig.~\ref{distributions}.
We can also see that the two discrete regions 
correspond to the right-sign limit ($\tan\beta \cos(\beta-\alpha) \simeq 0$) and wrong-sign limit ($\tan\beta \cos(\beta-\alpha) \simeq 2$)
as described around Eq. (\ref{xitau}).

\item The exotic Higgs decay $h\to AA$ or $h \to A \tau\tau$ is a promising channel
to probe the L2HDM explanation of the muon {$g-2$} as its branching ratio can be quite sizable
unless there is a particular reason to suppress $\lambda_{hAA}$ as shown in Fig.~\ref{EXOdecay}.

\end{itemize}

\section{$\tau-$rich signature at LHC} 
\label{sec:LHC}

In the previous section, we identified two favored regions of the L2HDM parameter space.
In this section, we discuss how the current LHC search results can
constrain this model further. Since the relationship between $m_A$ and $\tan\beta$
is constrained by the $\gmtwo$, as shown in the left panel of Fig.~\ref{distributions},
we can simply parametrize $\tan\beta$ as a function of $m_A$:
\begin{equation}
\tan\beta=1.25 \left(\frac{m_A}{\gev}\right)+ 25,
\end{equation}
which will be assumed in this section. We left with three Higgs mass parameters 
$m_A, m_H$, and  $m_{H^\pm}$, which determine phenomenologies at the LHC.

The bulk parameter space with $m_A \ll m_H \sim m_{H^\pm}$ is a clear prediction of the lepton-specific 2HDM considered in this paper. Since the extra Higgs bosons are mainly from 
the ``leptonic'' Higgs doublet with a large $ \tan \beta$, all the three members are expected to dominantly decay into the $\tau-$flavor, leading to $\tau-$rich signatures at LHC~\cite{Su:2009fz,Kanemura:2011kx,Kanemura:2014bqa} via the following production and ensuing cascade decay chains:
\begin{align}
pp\ra& W^{\pm *} \to H^\pm A \to (\tau^\pm \nu)(\tau^+\tau^-),\\
pp\ra  &       Z^*/\gamma^* \to HA \to  (\tau^+\tau^-)(\tau^+\tau^-),\\
pp\ra& W^{\pm *} \to H^\pm H \to (\tau^\pm \nu)(\tau^+\tau^-),\\
pp\ra& Z^*/\gamma^* \to H^+H^- \to (\tau^+\nu)(\tau^-\bar\nu).
\end{align}
As seen in Fig.~\ref{degeneracy}, we can also find a small island at the right-lower corner of the plot
where $m_{H^\pm} \sim  m_A \sim 100$~GeV, which we call Region B while the above bulk region 
we call Region A. In the following we fix $m_{H^\pm}$ in the two regions based on the best fit point:
\begin{itemize}
\item[]Region A: $m_{H^\pm} = m_{H} + 15$ GeV
\item[]Region B: $m_{H^\pm} = \max (90$ GeV, $0.8m_{A} + 10$ GeV)
\end{itemize}
With these relations we explore $m_A$-$m_H$ plane.

\begin{figure}[htb]
\includegraphics[width=3.1in]{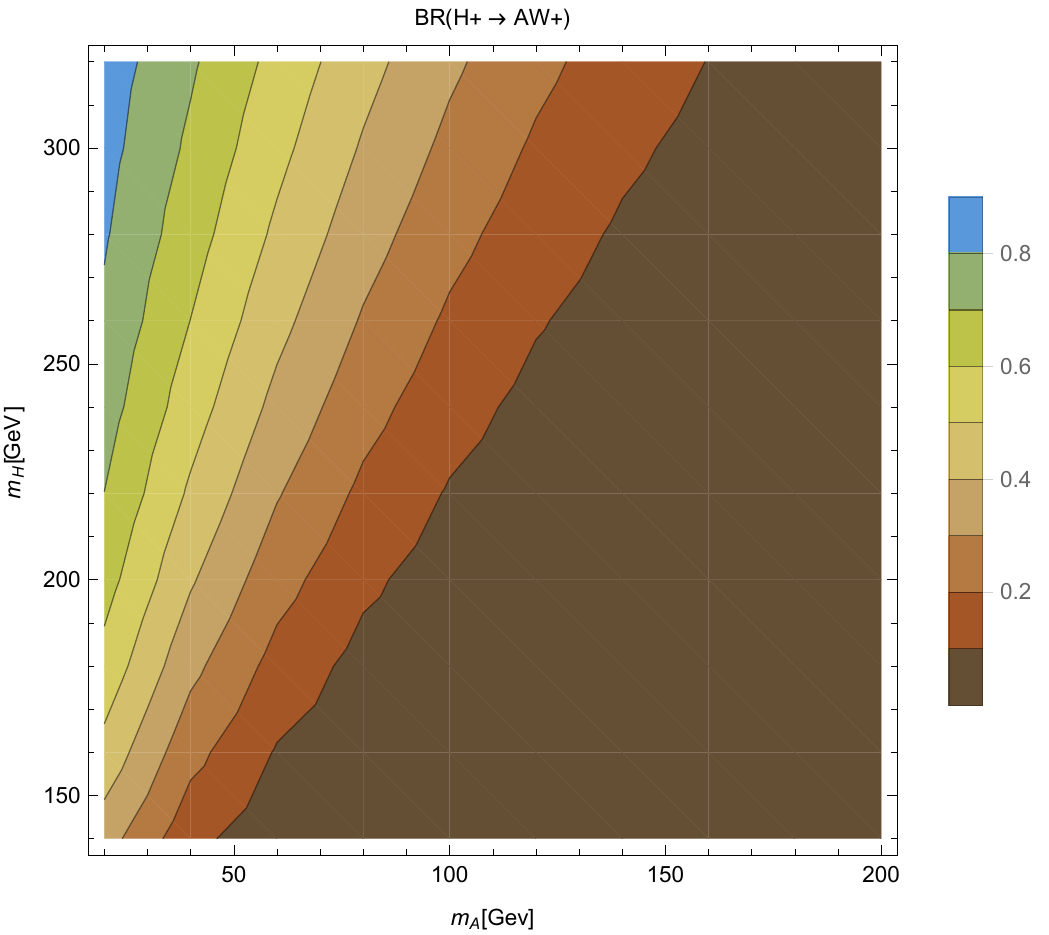}
\includegraphics[width=3.1in]{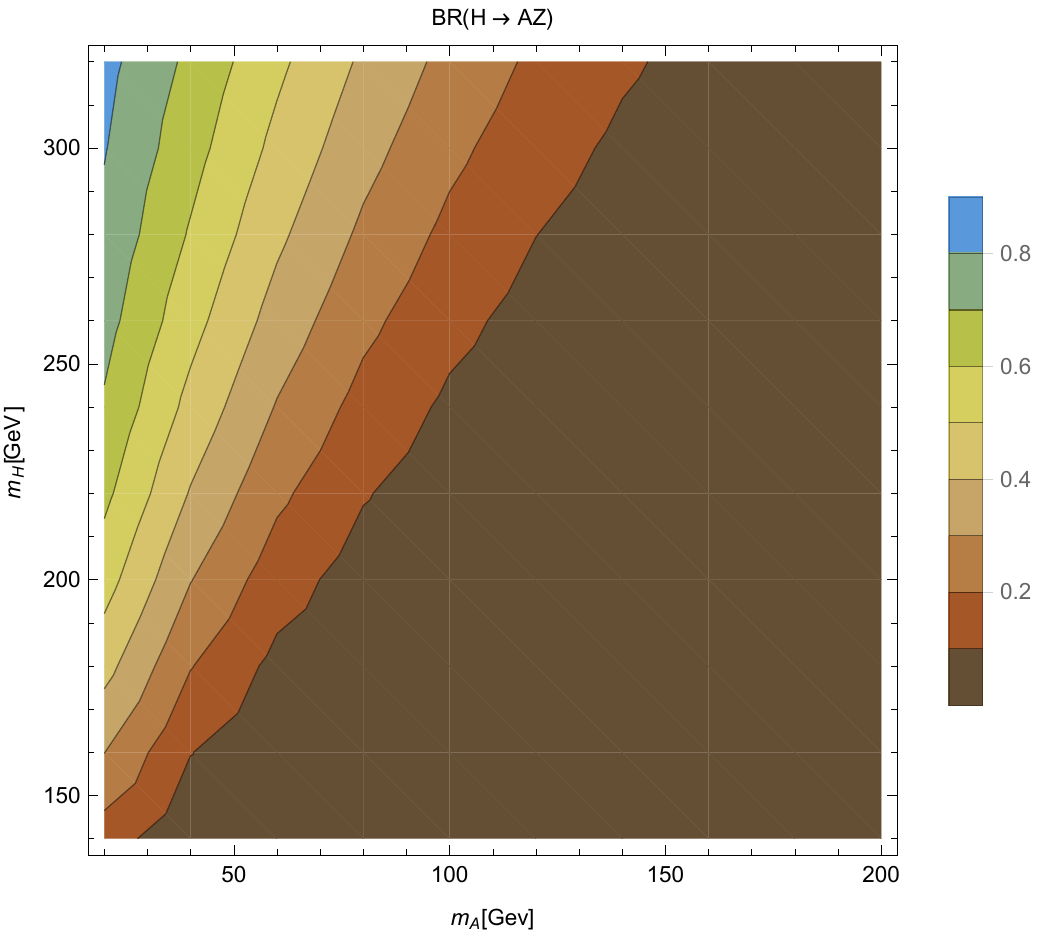}
\caption{Contour plot of branching ratio Br$(H^+\to A W^+$) and Br$(H \to A Z$).
Br$(H^+\to A W^+$) + Br$(H^+\ra \tau^+\nu$) $\simeq 1$ in Region A. The relation
$\tan\beta=1.25 m_A + 25$ is used.}
\label{HCdecay}
\end{figure}

A large $\tan\beta$ enhances the lepton Yukawa couplings of extra Higgses $H^+/H/A$ leading to a fast decay into tau leptons in general. The light pseudo-scalar $A$ indeed decays into $\tau \tau$ essentially at 100\%, however, the heavier $H^\pm/H$, in the presence of this light $A$, can sizably decay into $A W^\pm/Z$ via electroweak gauge interactions. This partial decay width is enhanced by the well-known factor $(m_{H^+/H}^2/M_W^2)^2$ in the limit $m_{H^+/H}^2\gg M_{W/Z}^2$ and expressed as
\begin{align}\label{WA}
\Gamma(H^+\ra W^+ A)=&\f{1}{16\pi}\f{M_W^4}{v^2m_{H^+}}\ld(1,m_{H^+}^2/M_W^2,m_A^2/M_W^2)\ld^{1/2}(1,M_W^2/m_{H^+}^2,m_A^2/m_{H^+}^2)\cr
\ra& \f{1}{16\pi}\L\f{m_{H^+}}{v}\R^2m_{H^+}\quad {\rm for}~ m_{H^+}^2\gg M_W^2,
\end{align}
where   $\ld(1,x,y)= (1-x-y)^2 -4 xy$. It can be compared with the partial decay width of $H^+ \to \tau\nu$
\begin{align}\label{taunu}
\Gamma(H^+\ra \tau^+ \nu)=&\f{m_{H^+}}{16\pi}\L \f{\sqrt{2}m_\tau}{v}\tan\beta \R^2.
\end{align}
From Eqs.~(\ref{WA})  and (\ref{taunu}) one can see that the $WA$ channel turns out to dominate over the $\tau\nu$ channel when $m_{H^+}>\sqrt{2}m_\tau \tan\beta$ as shown in the left panel of Fig.~\ref{HCdecay}, where we plotted the branching ratio of $H^\pm \to A  W^\pm$.
We can get the decay width $\Gamma(H\ra ZA)$ by replacing $m_{H^+}$ and $M_W$ with $m_H$ and $M_Z$, respectively, in the above expression, and its branching ratio is also shown in the right panel.

Even if $H/H^\pm$  undergoes the decay involving $Z/W^\pm$, the associated $A$ will eventually decay into $\tau\tau$ and thus multiple $\tau$ signature up to $4\tau + W$ or/and $Z$ would be one of the peculiar signatures
of the model at the LHC.

\begin{figure}[htb]
\includegraphics[width=3.1in]{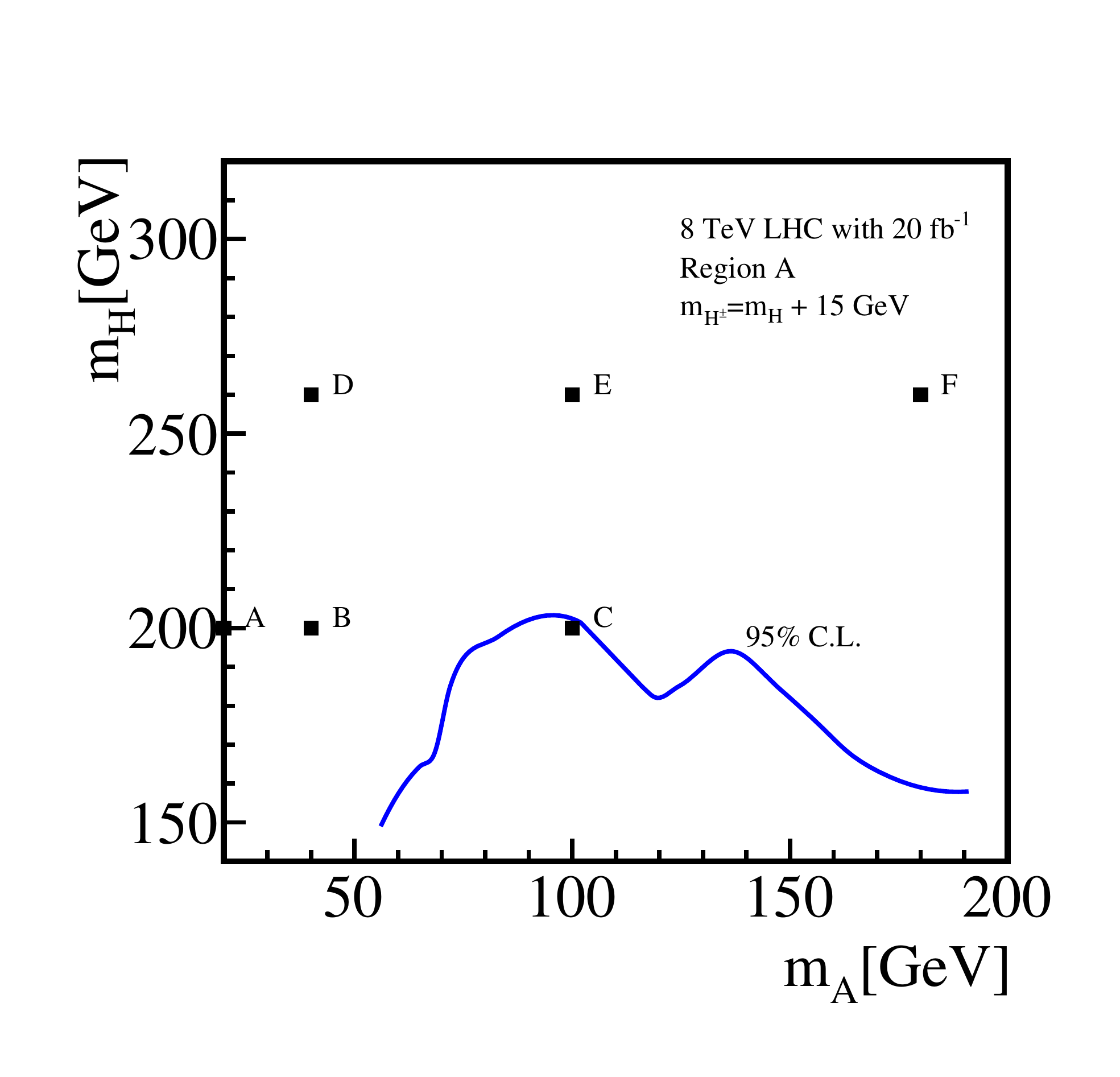}
\includegraphics[width=3.1in]{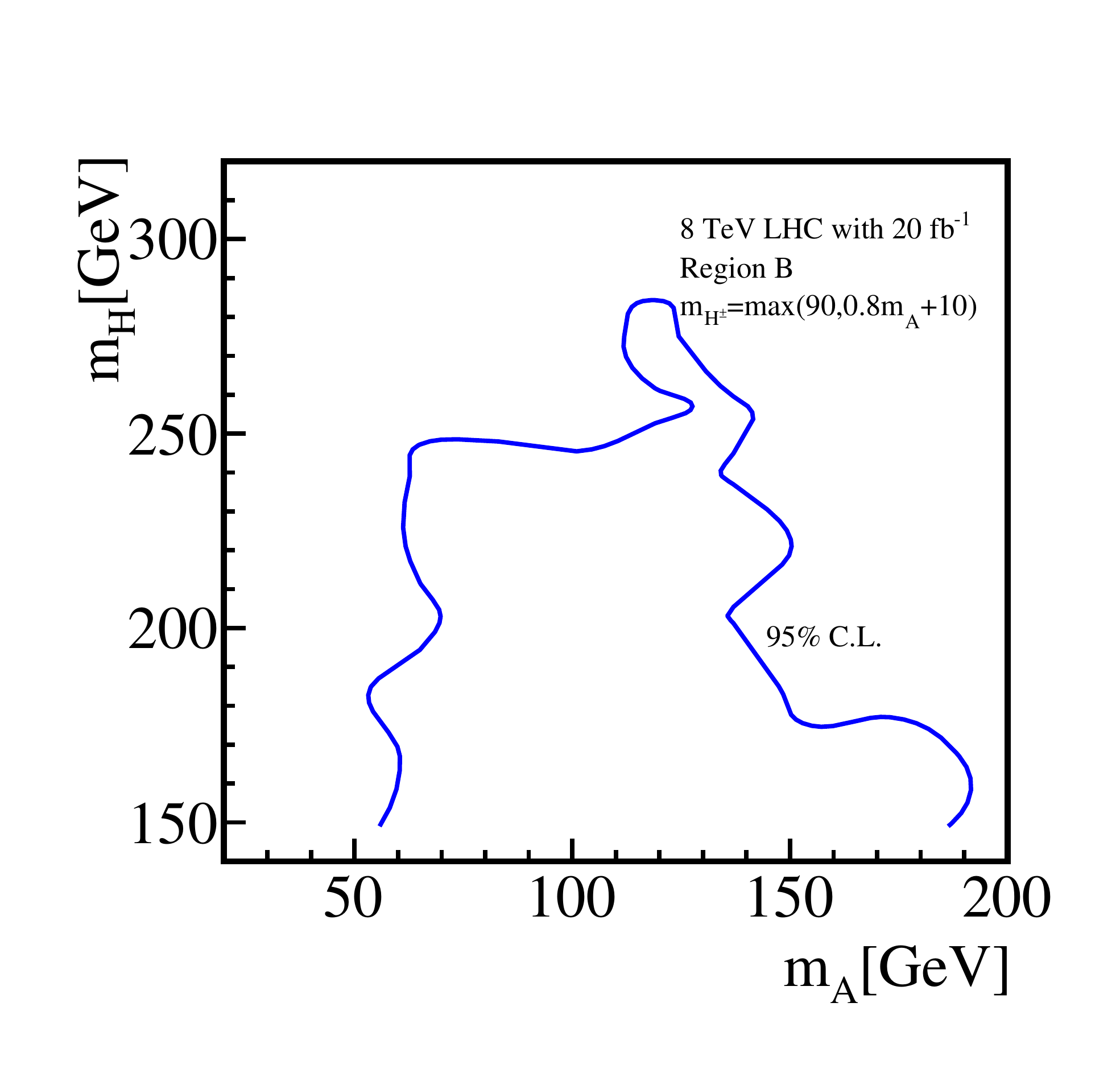}
\caption{95\% CL. contour from the chargino-neutlarino search at LHC 8TeV shown in $m_A $ vs. $m_H$ plane for Region A (left) and Region B (right).}
\label{Limitlhc8}
\end{figure}

\subsection{Current Constraints}

Current LHC 8 TeV data already set the constraints in the parameter space we are interested in.
In both Region A and Region B we take model point grid with 
$m_A \in [20, 200]$~GeV and $m_H \in [140, 320]$~GeV
both with 20~GeV steps, that is, 100 model points for each region.
We generate the 50,000 signal events with \texttt{MadGraph}~\cite{madgraph} for each parameter point and interfaced to \texttt{CheckMATE 1.2.0-beta}~\cite{Checkmate}  for checking the current bound with 
20 fb$^{-1}$ data at 8 TeV LHC.
The analyses implemented in the \texttt{CheckMATE} are listed in the Table~\ref{tablecheckmate}. 
\textit{We checked all the analyses and considered that a model point is excluded 
when at least one analysis excludes it at 95\% C.L.}

Fig.~\ref{Limitlhc8} shows the estimated 95\%~C.L. exclusion contours. 
For most of the parameter space, the strongest constraint comes
from the chargino-neutralino search in ATLAS~\cite{Aad:2014nua}. 
Especially, it is from the signal region ``SR2$\tau$a" therein, which requires 
two $\tau$ leptons and an additional isolated lepton, with $m_{T2}^{\max} > 100$~GeV, 
$E\!\!\!/_T>50$~GeV and $b$-veto. 
Heavier $m_H > 200$~GeV (Region A) or $m_H > 280$~GeV (Region B), and light $m_A<50$~GeV
are still allowed and we will show later that the next run of LHC can explore 
some of the regions. For the heavier $m_H$ regions the sensitivities are weaker just because of the smaller cross sections, while for light $m_A$ region it is because $\tau$s from lighter $A$ decays 
become softer and thus the acceptance quickly decreases. Moreover, the 
$H/H^\pm \to A Z/W^\pm$ decay modes also 
start open to decrease the number of hard $\tau$s from direct $H/H^\pm$ decays.
Note that the exclusion of the lighter $m_A$ parameter space is of interest only for Region A, since for Region B the interesting $m_A$ in our scenario to explain $(g-2)_\mu$ is confined to be lie above $100$~GeV as you can see in Fig.~\ref{degeneracy}.

\begin{table}[htb]
{\small
\begin{tabular}{llr}
\hline
arXiv number & description & integrated luminosity [fb$^{-1}$]\cr
\hline
atlas-1308-2631       & ATLAS, 0 leptons + 2 b-jets + etmiss      &        20.1 \cr
atlas-1402-7029       & ATLAS, 3 leptons + etmiss (chargino+neutralino)  & 20.3  \cr
atlas-1403-4853       & ATLAS, 2 leptons + etmiss (direct stop)        &   20.3   \cr
atlas-1403-5294       & ATLAS, 2 leptons + etmiss, (SUSY electroweak)   &  20.3  \cr
atlas-1403-5294-CR  &     ATLAS, 2 leptons + etmiss CR, (SUSY electroweak) & 20.3  \cr
atlas-1404-2500       &   ATLAS, Same sign dilepton or 3l               &    20.3   \cr
atlas-1407-0583       &   ATLAS, 1 lepton + (b-)jets + etmiss (stop)   &     20.3   \cr
atlas-1407-0600       &    ATLAS, 3 b-jets + 0-1 lepton + etmiss           &  20.1  \cr
atlas-1407-0608       &    ATLAS, Monojet or charm jet (stop)               & 20.3   \cr
atlas-1502-01518      &    ATLAS, Monojet plus missing energy             &   20.3   \cr
atlas-conf-2012-104   &    ATLAS, 1 lepton + $\ge 4$ jets + etmiss             & 5.8    \cr
atlas-conf-2012-147   &    ATLAS, Monojet + etmiss                          & 10.0   \cr
atlas-conf-2013-021   &    ATLAS, WZ standard model (3 leptons + etmiss)   &  13.0   \cr
atlas-conf-2013-024   &    ATLAS, 0 leptons + 6 (2 b-)jets + etmiss        &  20.5   \cr
atlas-conf-2013-031   &    ATLAS: Higgs spin measurement (WW)           &     20.7   \cr 
atlas-conf-2013-036   &    ATLAS: 4 leptons + etmiss                       &  20.7   \cr
atlas-conf-2013-047   &   ATLAS, 0 leptons + 2-6 jets + etmiss          &    20.3   \cr  
atlas-conf-2013-049   &    ATLAS, 2 leptons + etmiss                       &  20.3   \cr
atlas-conf-2013-061   &    ATLAS, 0-1 leptons + $\ge 3$ b-jets + etmiss    &     20.1   \cr 
atlas-conf-2013-062   &   ATLAS: 1-2 leptons + 3-6 jets + etmiss           & 20.1   \cr
atlas-conf-2013-089   &   ATLAS, 2 leptons (razor)                       &   20.3   \cr
atlas-conf-2014-014   &    ATLAS, 2 leptons + b-jets (stop)             &     20.3   \cr 
atlas-conf-2014-033   &    ATLAS, WW standard model measurement        &      20.3   \cr  
atlas-conf-2014-056   &    ATLAS, ttbar spin correlation measurement       &  20.3   \cr
cms-1303-2985         &   CMS, alpha-T + b-jets                          &   11.7   \cr  
cms-1301-4698-WW     &    CMS, WW standard model measurement             &   3.5  \cr
cms-1405-7570         &  CMS, Various chargino and neutralino           &   19.5   \cr
cms-smp-12-006       &    CMS, WZ standard model (3 leptons + etmiss)    &   19.6   \cr
cms-sus-12-019        &   CMS, 2 leptons, $\ge$ 2 jets + etmiss (dilep edge)  & 19.4   \cr 
cms-sus-13-016        &    CMS, OS lep 3+ b-tags                           &  19.5  \cr
\hline
\end{tabular}
\caption{The list of the analysis used in our analysis implemented in the \texttt{CheckMATE}.
The list is found in the \texttt{CheckMATE/data/} directory.} 
}
\label{tablecheckmate}
\end{table}

\subsection{14 TeV prospects}
\begin{figure}[htb]
\includegraphics[width=2.12in]{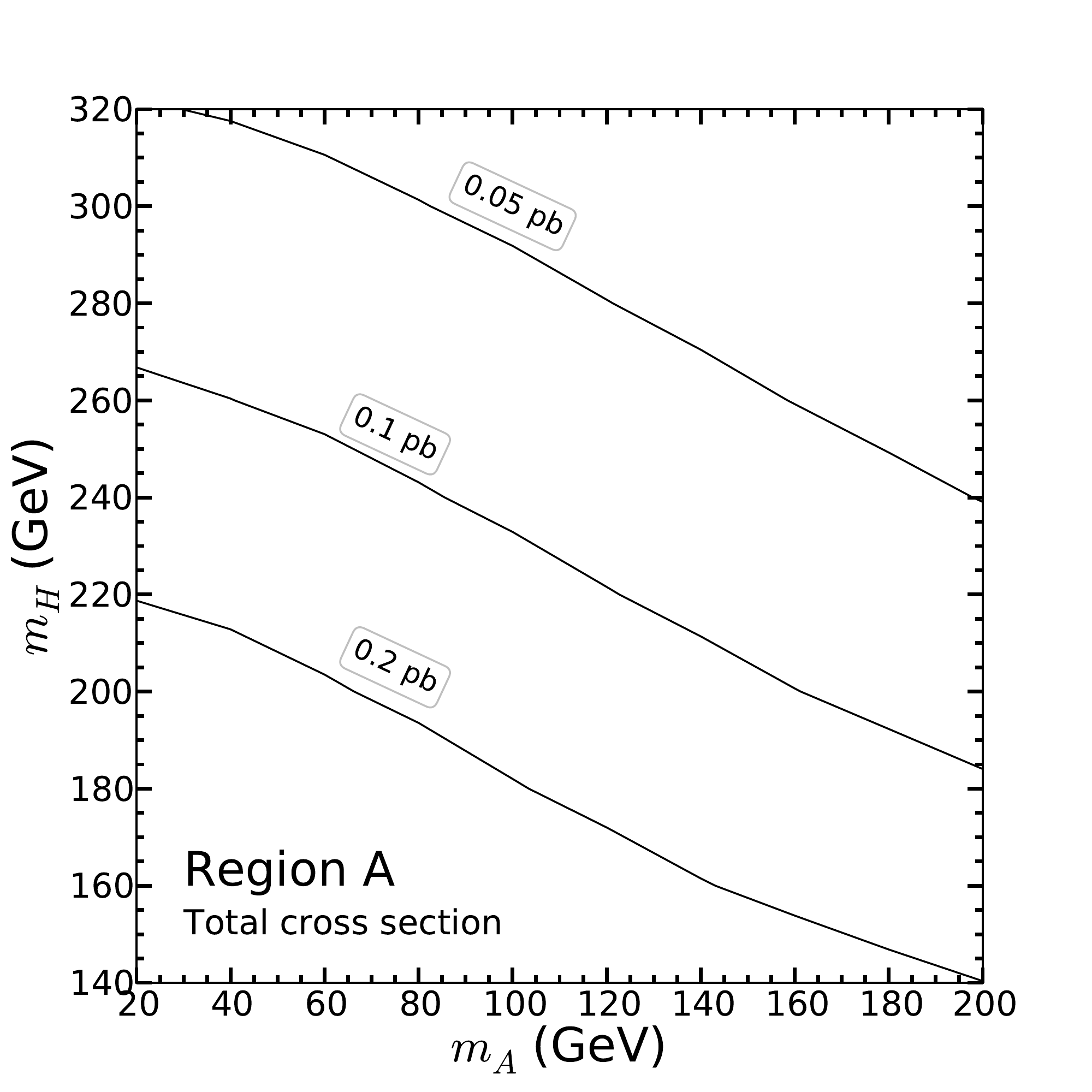}
\includegraphics[width=2.12in]{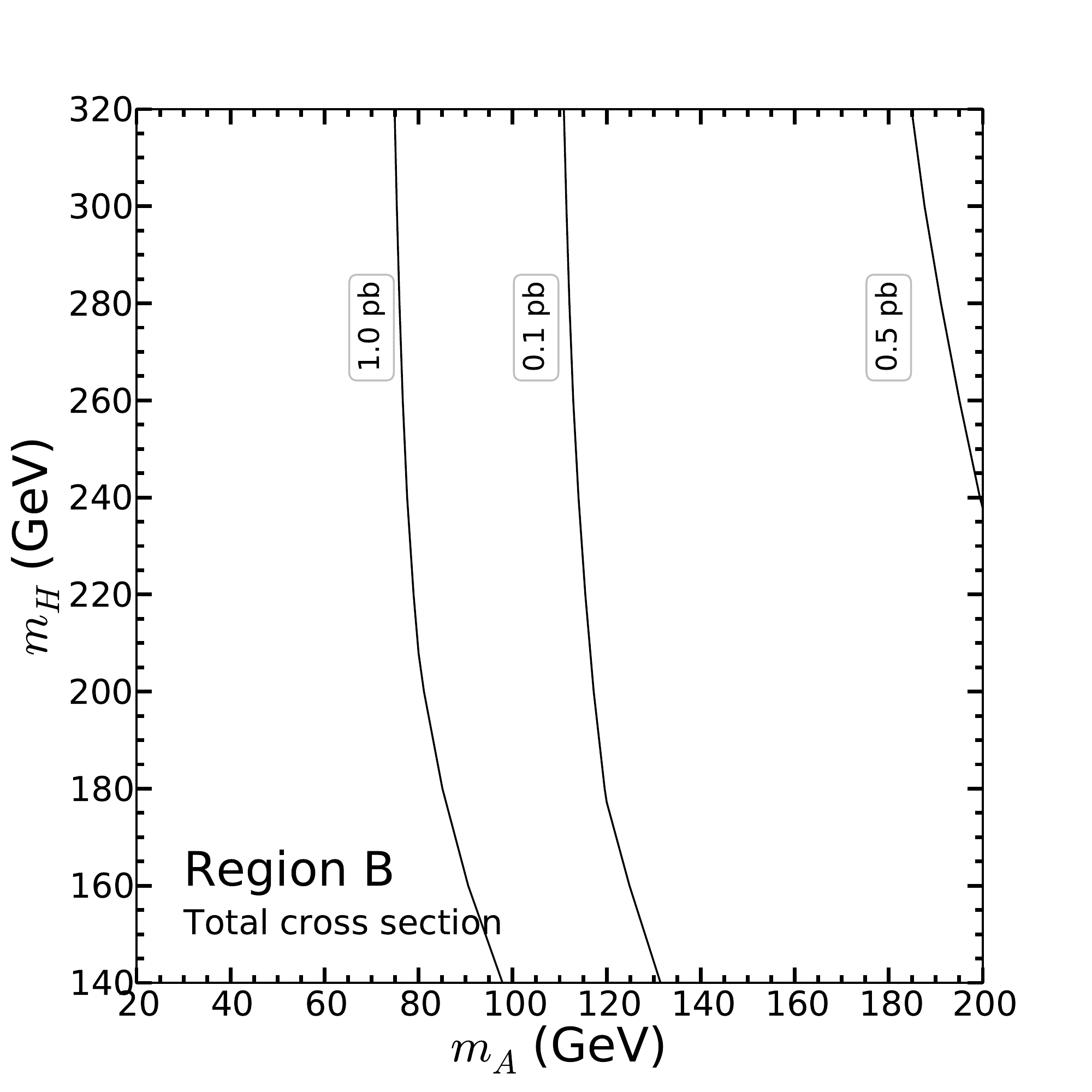}
\includegraphics[width=2.12in]{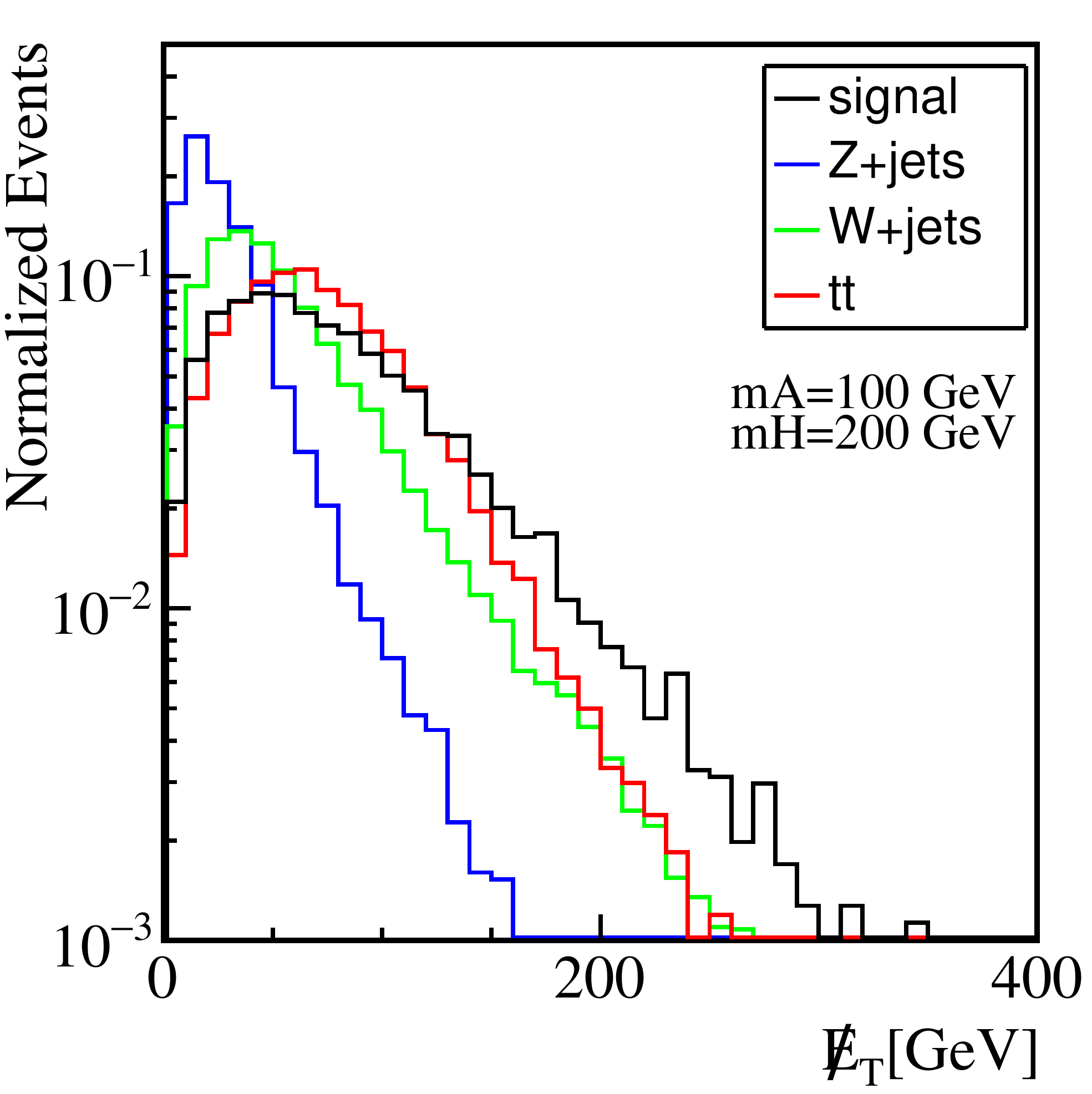}
\caption{Total signal cross section dependence in $m_A$ vs. $m_H$ plane in Region A (left) 
and Region B (center). 
Right panel: Missing transverse momentum distributions for the signal benchmark point~C ($m_A=100$~GeV and $m_H=200$~GeV in Region A) and various BG processes. }
\label{fig:crosssections}
\end{figure}

In this section we estimate the reach of the LHC 14 TeV in Region A and B 
based on the model point grids defined previously for the LHC 8 TeV study.
The signal cross sections depend on heavy Higgs masses, and in Fig.~\ref{fig:crosssections} we show the contour plots of total cross section on the $m_A-m_H$ plane for Region A (Region B) in the left (center) panel. Actually, for relatively small $m_A$ the dominant contribution comes from the $H^\pm A$ production while the 
$HA$ production contributes secondarily; $HH^\pm$ and $H^+H^-$ contributions are subdominant.

For the Standard Model background processes
we consider $t\bar{t}$, $W$+jets, $Z$+jets, and di-boson 
productions $W^+W^-, W^\pm Z, ZZ$. All background events are generated with 
 \texttt{ALPGEN}~\cite{alpgen} +  \texttt{Pythia}~\cite{pythia}.
We only consider leptonic decay modes including tau for all processes as
later on we select events with at least 3 leptons including taus.
To include the mis-tagging-$\tau$ effects, we generate 
the MLM-matched samples~\cite{mlm} with 2 to 3 additional jets
for $W$+jets, and with 1 to 2 additional jets for $Z$+jets.
Cross sections with the above generation cut are 
$102$~pb, $1365$~pb, $714$~pb, $8.13$~pb, 
$0.942$~pb and $0.112$~pb for $t\bar{t}$, $W$+jets, $Z$+jets, $W^+W^-$, 
$W^\pm Z$, and $ZZ$, respectively.

As this model predicts $\tau$-rich signatures the signal is sensitive to $\tau$-tagging and 
we implement $\tau$-tagging algorithm using track and calorimeter information from \texttt{Delphes 3.0}~\cite{delphes3}
as described in 
Ref.~\cite{Papaefstathiou:2014oja}, which basically is a simplified version of the 
ATLAS $\tau$-tagging algorithm~\cite{Aad:2014rga,TheATLAScollaboration:2013wha}.
We use two variables: 
\begin{eqnarray}
 \ \  R_{\max} = \max_{\rm tracks} \Delta R(p_j, p_{i})
 \ \ {\rm and} \ \  f_{\rm core}=\frac{\sum_{R<0.1} E^{\rm calo}_T}{\sum_{R<0.2} E^{\rm calo}_T},
\end{eqnarray}
where $p_j$ is the jet center direction and the distance of the furthest track among $p_i$ (with $p_T> 1$~GeV)  to $p_j$ is denoted as $R_{\max}$; $E_T^{\rm calo}$ is the $E_T$ deposited in each calorimeter tower and 
the summations run over the calorimeter towers within the cones centered around $p_j$ with cone size $R<0.1$ and $0.2$ for the numerator and the denominator, respectively. 
Both $R_{\max}$ and $f_{\rm core}$ measure essentially how narrow the jet is; $\tau$-jet is expected to be narrow and gives a smaller $R_{\max}$ and $f_{\rm core} \sim 1$. 
We found these two variables are most relevant for the discrimination.

\begin{figure}[htb]
\includegraphics[width=2.12in]{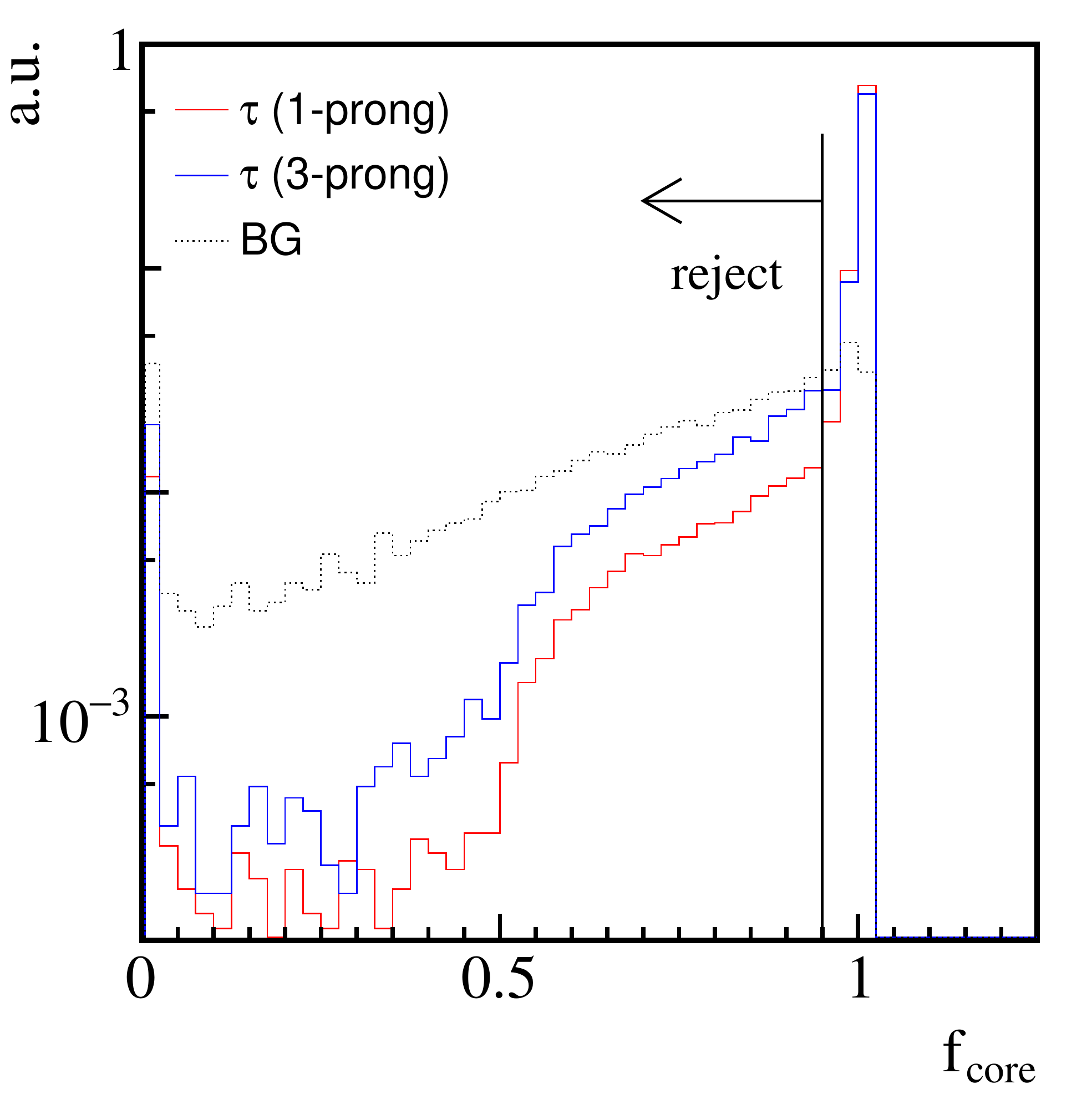}
\includegraphics[width=2.12in]{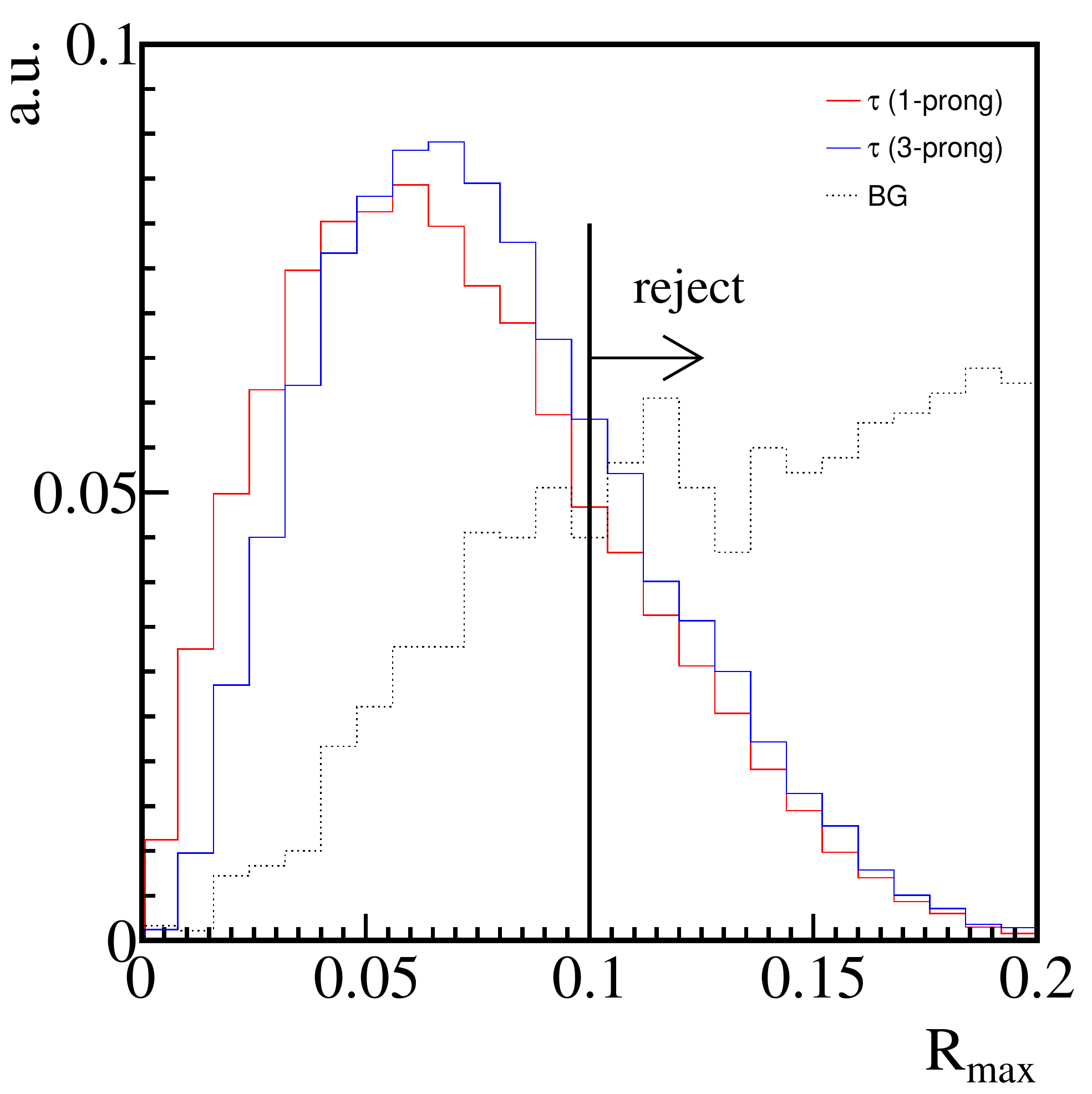}
\includegraphics[width=2.12in]{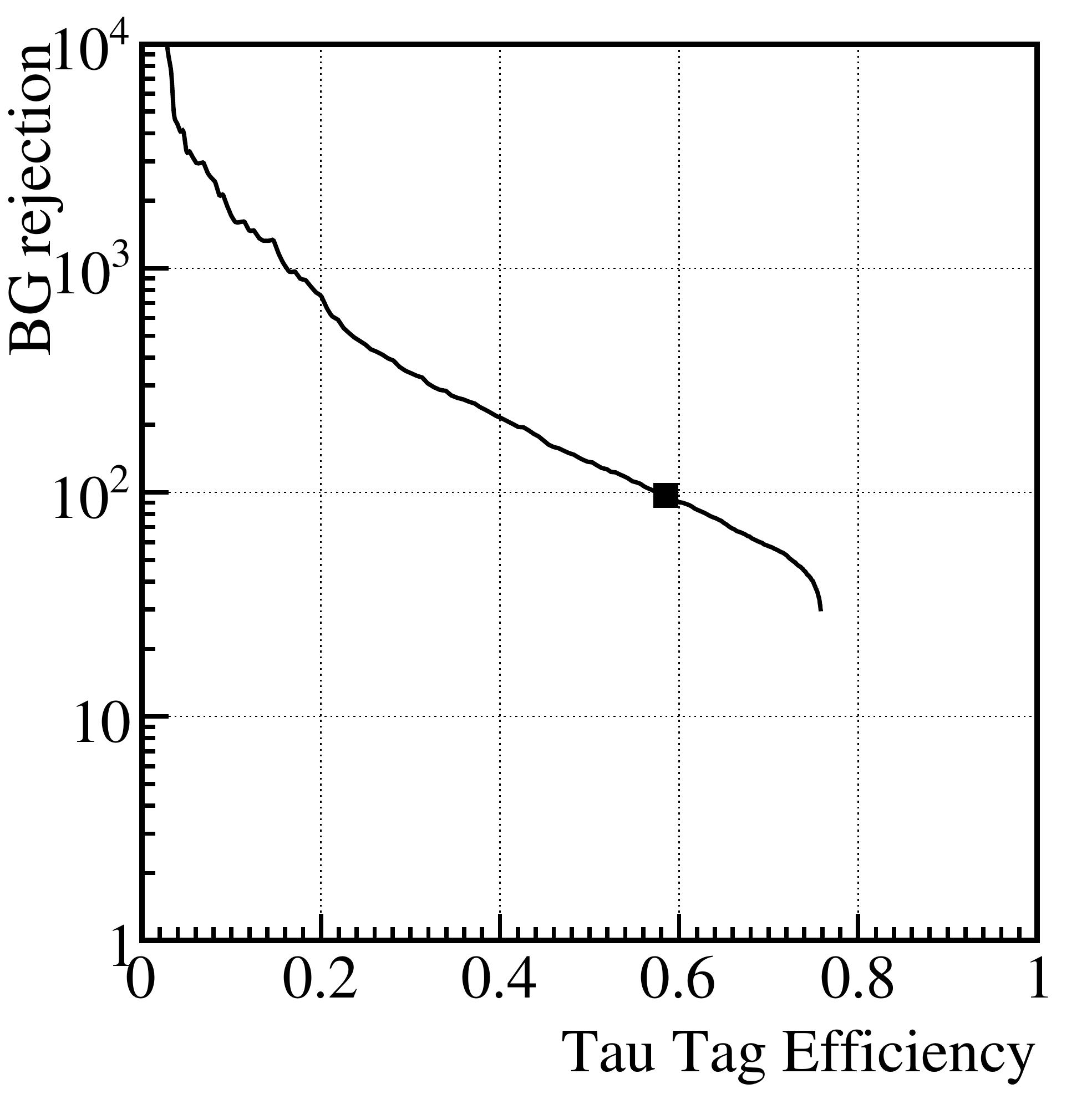}
\caption{ROC curve for our $\tau$-tagging algorithm. Our working point is denoted with a filled square, where $59\%$ efficiency 
with 1\% mis-identification efficiency for QCD jets is obtained.}
\label{tautag}
\end{figure}

We show  $R_{\max}$ and $f_{\rm core}$ distribution in Fig.~\ref{tautag}. 
We also show the ROC curve obtained by changing the cut value $R_{\max}^{\rm cut}$ for $R_{\max} < R_{\max}^{\rm cut}$ 
with fixing $f_{\rm core}^{\rm cut}=0.95$ for $f_{\rm core} > f_{\rm core}^{\rm cut}$.
Compared with the plot shown in Ref.~\cite{TheATLAScollaboration:2013wha}, our simulation is reasonably conservative up to the signal efficiency $\sim 60$\%.
We select the working point with $R_{\max}^{\rm cut} = 0.1$, which gives $\epsilon_\tau=59\%$ with the background jet rejection $1/\epsilon_{BG}=97$.

We apply the following event selection cuts to the signal and BG events.
First, we require events with at least 3 $\tau$-tagged jets, based on the algorithm explained above.
At this stage the dominant background becomes $t\bar{t}$, $W$+jets and $Z$+jets. 
Next, we require enough missing momentum $E\!\!\!/_T > 100$~GeV to efficiently reduce the $W$+jets and $Z$+jets contributions.
The normalized  $E\!\!\!/_T$ distributions are shown in the right panel of Fig.~\ref{fig:crosssections}. 
Finally, to reduces the $t\bar{t}$ background, we veto events with any $b$-tagged jet with $p_T>25$~GeV nor any jet with $p_T > 50$~GeV.
This cut efficiently reduces the remaining backgrounds.
Table~\ref{tab:selectioncut1} summarizes the number of events  
after the successive selection cuts in unit of fb for the various BG processes and for the signal
benchmark model point C. 
We compute the signal to background ratio $S/B$ and 
significance based on statistical uncertainty $S/\sqrt{B}$.
The significance quoted here is based on the integrated luminosity of 25 fb$^{-1}$. 
We can use the $\mu\mu$ modes as suggested in 
Ref.~\cite{Kanemura:2011kx} to improve the sensitivity and to reconstruct the events
but we mainly focus on $\tau$-rich signatures, which require a relatively low statistics to set limit
and expected sensitive at the early stage of LHC run 2.

We show the results for several selected benchmark points A to F in detail. 
Table~\ref{tab:selectioncut2} collects the numbers and significances including the other benchmark model points.

\begin{table}[htb]
\begin{tabular}{l|r|rrrrrr|r||rr}
selection cuts & point C & $t\bar{t}$ & $W$+jets & $Z$+jets & $WW$ & $WZ$ & $ZZ$ & total BG 
& $S/B$ & $S/\sqrt{B}_{25{\rm fb}^{-1}}$ \cr
\hline
\hline
total $\sigma_{\rm gen}$ [fb]  & 153.580 & $102  \cdot 10^3$ & $1365\cdot 10^3 $ & $714 \cdot 10^3 $ & 8125 & 942 & 112 & $2190 \cdot 10^3$ & - & - \cr
\hline
$n_{\ell} \ge 3$  & 21.713 & 273.27 & 138.59 & 3412.84 & 6.495 & 88.937 & 26.965 & 3947.1 & - & 1.7\cr
$n_{\tau} \ge 3$  & 4.386 & 5.837 & 13.776 & 91.324 & 0.070 & 0.343 & 0.174 & 111.52 & 0.04 & 2.1\cr
$E\!\!\!/_T > 100$~GeV   & 1.179 & 1.482 & 0.232 & 1.244 & 0.000 & 0.018 & 0.003 & 2.980 & 0.4 & 3.4\cr
$n_b=n_{j}=0$ & 0.857 & 0.163 & 0.000 & 0.505 & 0.000 & 0.017 & 0.003 & 0.688 & 1.2 & 5.2\cr
\hline
\end{tabular}
\caption{The number of events after applying successive cut for 14 TeV LHC. 
Benchmark point C ($m_A=100$~GeV, $m_H=200$~GeV) is shown for the signal.
The significance quoted  is based on integrated luminosity of 25 fb$^{-1}$.} 
\label{tab:selectioncut1}
\end{table}

\begin{table}[htb]
\begin{tabular}{l|rrrrrr}
 & point A & point B & point C & point D & point E & point F \cr
\hline
$m_A$~[GeV] & 20 & 40 & 100 & 40 & 100 & 180  \cr
$m_H$~[GeV] & 200 & 200 & 200 & 260 & 260 & 260  \cr
\hline
\hline
total $\sigma_{\rm gen}$ [fb] & 270.980 & 241.830 & 153.580 & 100.430 & 71.271 & 44.163\cr
\hline
$n_{\ell} \ge 3$  & 6.606 & 16.681 & 21.713 & 7.110 & 11.962 & 8.822\cr
$n_{\tau} \ge 3$  & 0.894 & 2.602& 4.386 & 0.888 & 2.346 & 1.971\cr
$E\!\!\!/_T > 100$~GeV    & 0.201 & 0.547 & 1.179 & 0.209 & 0.765 & 0.926\cr
$n_b=n_{j}=0$& 0.098 & 0.314  & 0.857 & 0.121  & 0.479 & 0.631\cr
\hline
\hline
$S/B$ & 0.1  & 0.5 & 1.2 & 0.2 &0.7 & 0.9 \cr
$S/\sqrt{B}_{25{\rm fb}^{-1}}$ & 0.6 & 1.9 & 5.2 & 0.7 & 2.9 & 3.8\cr
\hline
\end{tabular}
\caption{The number of events after applying successive cut for 14 TeV LHC. The significance quoted  is based on integrated luminosity of 25 fb$^{-1}$.} 
\label{tab:selectioncut2}
\end{table}

\begin{figure}[htb]
\includegraphics[width=3.21in]{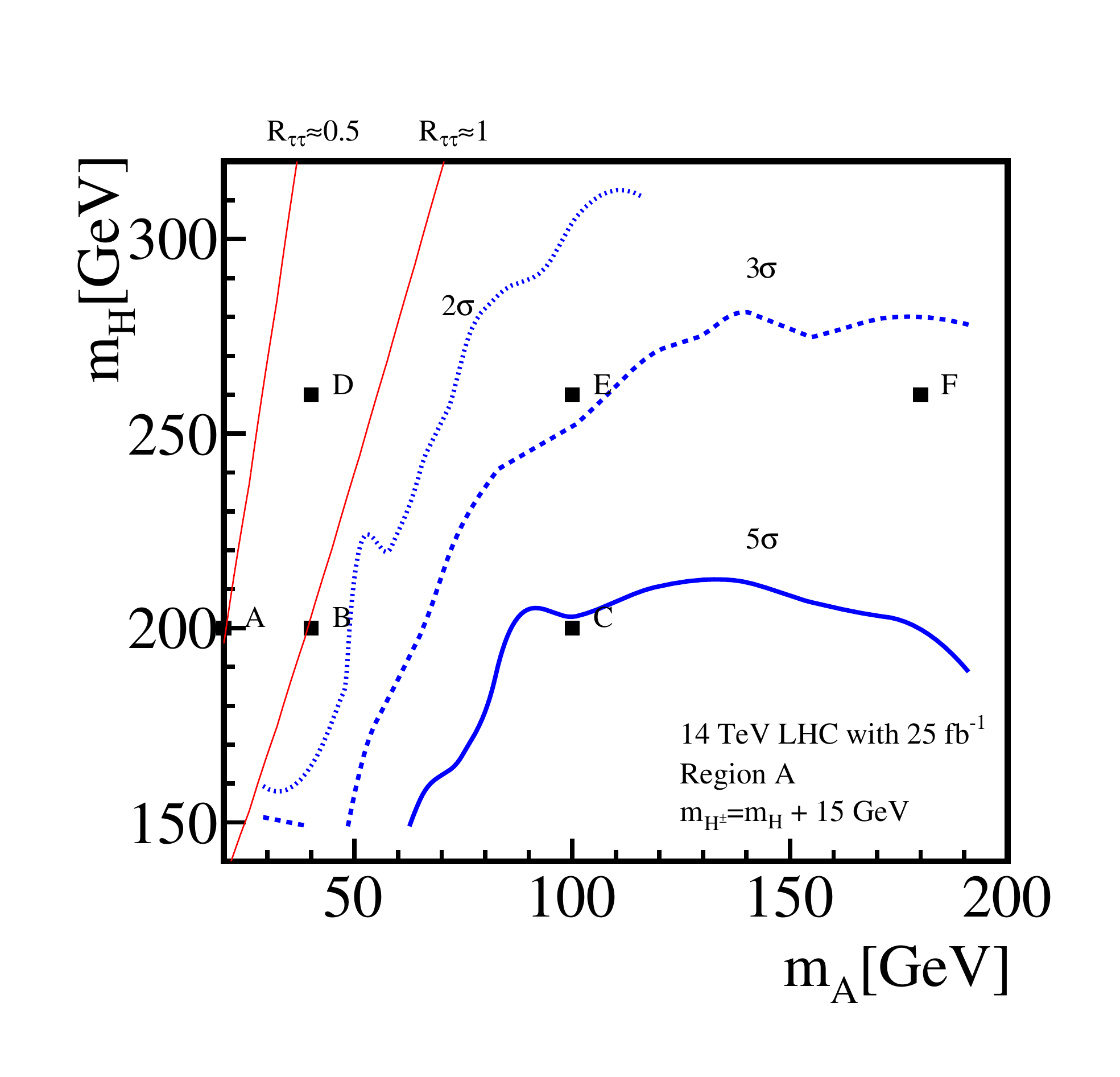}
\includegraphics[width=3.21in]{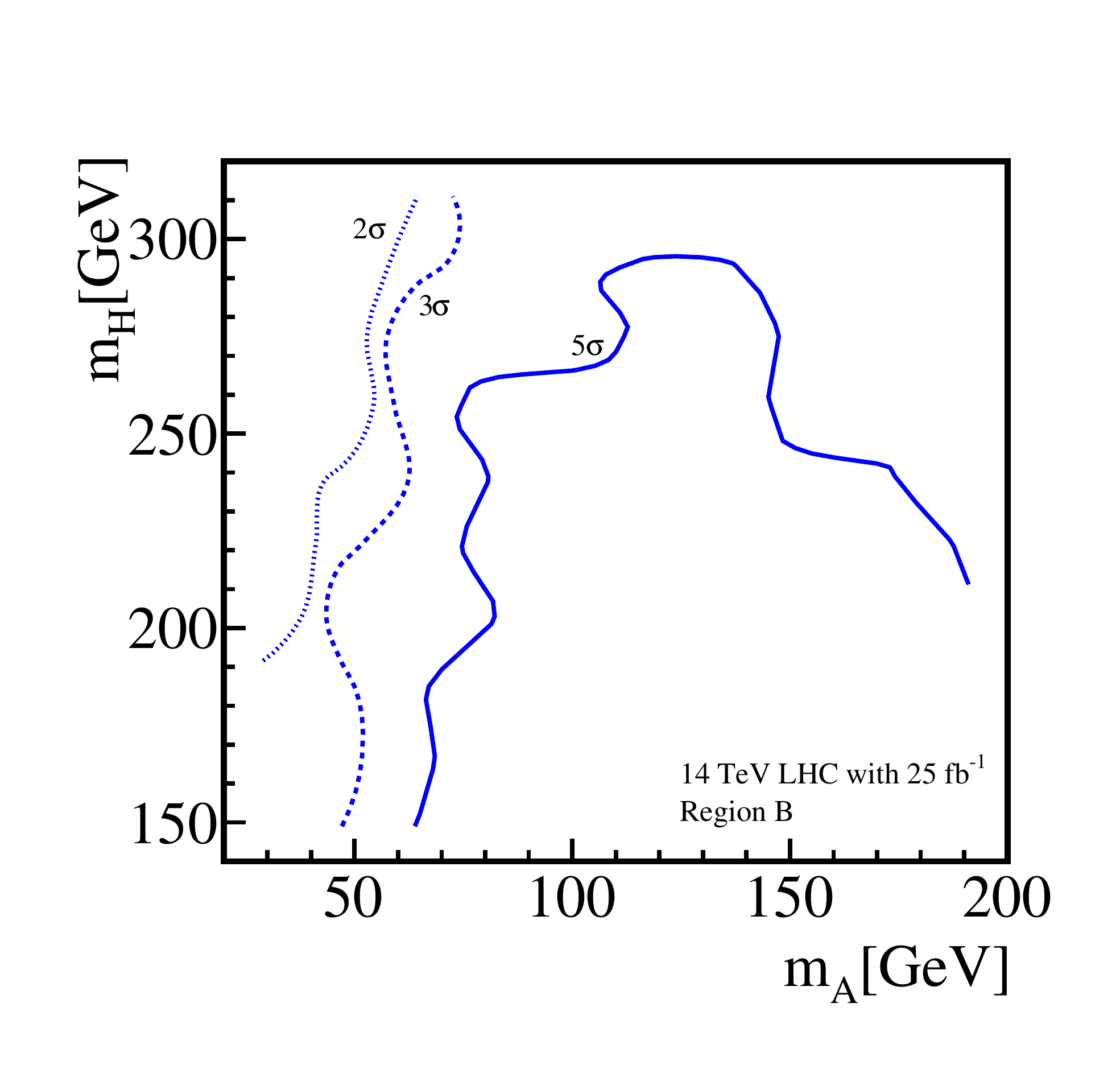}
\caption{
$2\sigma$, $3\sigma$ and $5\sigma$ discovery reach contours at LHC 14 TeV shown in $m_A $ vs. $m_H$ plane for Region A (left) and Region B (right). Assumed integrated luminosity here is 25 fb$^{-1}$.
Benchmark points selected in Table~\ref{tab:selectioncut2} are indicated with filled boxes.
Red lines indicate the region with expected smaller $\tau$ separation of $R_{\tau\tau} \sim 0.5$ and $1$.}
\label{reachLHC14}
\end{figure}

Based on the significance values we show the expected discovery reaches at LHC 14 TeV 
in Fig.~\ref{reachLHC14}. The left panel corresponds to Region A and the right panel does to Region B.
Both panels show the expected $2\sigma$, $3\sigma$ and $5\sigma$ discovery reach contours with assumed integrated luminosity of 25 fb$^{-1}$.
It is seen that most of the interesting parameter regions can be covered. Only limitation is for the region 
with light $m_A$ and heavy $m_H$ where the sensitivity becomes weak even though the intrinsic signal cross sections 
are not so small.
The reasons are again because of the smaller acceptance for the softer $\tau$ and longer 
decay chains involving $Z/W$ as explained in the previous section on 8~TeV analysis.
Moreover, in such a region, a light $A$ from heavy $H^+/H$ decay will be boosted,
resulting in a collimated $\tau-$pair which becomes difficult to be tagged as two separated $\tau$-jets.
It is one of the reasons to have less acceptance for this parameter region.
We can estimate the separation $R_{\tau\tau}$ of the $\tau$ leptons from $A$ decay:  
\begin{align}
R_{\tau\tau} \sim \frac{2m}{p_T} \sim \frac{4m_A}{m_{H^\pm/H}\sqrt{1 - 2\frac{m_A^2 + m_{W/Z}^2}{m_{H^\pm/H}^2} + \frac{(m_A^2 - m_{W/Z}^2)^2}{m_{H^\pm/H}^4} }}.
\end{align}
For example, $R_{\tau\tau} \sim 0.4$ for $m_H=300$~GeV and $m_A=30$~GeV, 
and $R_{\tau\tau} \sim 0.3$ for $m_H=400$~GeV and $m_A=30$~GeV. 
Since the 
jets are usually defined with $R=0.5$, the $\tau-$pair starts overlapping. 
We indicated the region with the overlapping $\tau$ problem in red lines in the left 
panel of Fig.~\ref{reachLHC14}. 
In that region, we have to think 
of how to capture the kinematic features of the boosted $A\ra\tau^+\tau^-$. 
We may be able to take the overlapping $\tau$ problem as an advantage by utilizing jet substructure study, which is already proven useful~\cite{Plehn:2010st,Altheimer:2012mn,Altheimer:2013yza}. 
For example, using di-tau tagging as proposed in Ref.~\cite{Katz:2010iq} might be beneficial,
although we leave this for future work.

\section{Conclusions}
\label{sec:conclusion}

The lepton-sepcific (or type X) 2HDM is an interesting option for the explanation of the muon {$g-2$} anomaly which requires a light CP-odd Higg boson $A$ and large $\tan\beta$. 
In this paper, we  made a scan of the L2HDM parameter space to identify the allowed ranges of the extra
Higgs boson masses as well as the related  two couplings $\xi_h^l$  and  $\lambda_{hAA}$ of the 125 GeV Higgs boson which govern its standard and exotic decays $h\to \tau^+\tau^-$ and $h\to AA/AA^*(\tau^+\tau^-)$, respectively.  The tau Yukawa coupling is found to be either in the wrong- or right-sign limit depending on the mass of $A$.  More precise determination of the standard tau Yukawa coupling
and a possible observation of one of the above exotic modes would provide a hint for the current scenario.

There appear two separate mass regions in favor of the muon {$g-2$}: (A) $m_A \ll m_{H} \sim m_{H^\pm}$ and (B) $m_A \sim m_{H^\pm} \sim 100 \mbox{GeV} \ll m_H$, which lead us to set up two regions of interest for the LHC study:  (A) $m_{H^\pm} = m_H + 15 \mbox{GeV}$, and (B) $m_{H^\pm} = \mbox{max}(90 \mbox{GeV}, 0.8 m_A + 10 \mbox{GeV})$ with $\tan\beta$ parametrized by $\tan\beta= 1.25 (m_A/\mbox{GeV})+25$.  In these parameter spaces, one expects to have $\tau$-rich signatures readily accessible at the LHC through the extra Higgs productions $pp \to AH^\pm/AH/H^\pm H^\pm /HH$ followed by $H \to A Z/ \tau^+\tau^-$ $H^\pm \to A W^\pm/\tau^+ \nu$ and $A \to \tau^+\tau^-$. Indeed,  the current LHC8 data start to exclude (yet mild) some of the above two regions:
$m_H$ up to about (A) 200 GeV and (B) 280 GeV for $m_A>50$ GeV from the consideration of the ATLAS neutralino-chargino search results. However, the region of $m_A \lesssim 30$ GeV (with $\tan\beta \lesssim 40$) which also satisfies the tau decay and lepton universality data \cite{abe15} is hardly tested by the $\tau$-rich signatures in near future even though 
HL-LHC should be able to over the region.
Thus, further study, for example, on the boosted $A\to \tau\tau$ will be required 
in the next runs of LHC to cover all of the L2HDM parameter space explaining the muon {$g-2$} anomaly.

\section*{Acknowledgment}

We would like to thank for helpful discussions with Lei Wang and the early collaboration with Daheng He. 
We initiated the idea of this paper at 2nd KIAS-NCTS Joint Workshop.
EJC is supported by the NRF grant funded by the Korea government (MSIP) (No. 2009- 0083526) through KNRC at Seoul National University. M.T. and Y.S.T. were supported by World Premier International Research
Center Initiative (WPI), MEXT, Japan.


\end{document}